\documentclass[12pt]{article}
\usepackage{cite,epsfig,amssymb,amsmath,graphicx,color}
\evensidemargin \oddsidemargin

\newcommand{\nn}{\nonumber}

\begin{document}

\begin{center}
{{{\Large \bf Gravity from Entanglement and RG Flow \\ in a Top-down Approach}
}\\[17mm]
O-Kab Kwon$^{1}$,~~Dongmin Jang$^{1}$,~~Yoonbai Kim$^{1}$,~~D.~D. Tolla$^{1,2}$\\[3mm]
{\it $^{1}$Department of Physics,~BK21 Physics Research Division,
~Institute of Basic Science, Sungkyunkwan University, Suwon 440-746, South Korea\\
$^{2}$University College,\\
Sungkyunkwan University, Suwon 440-746, South Korea}\\[2mm]
{\it okab@skku.edu,~dongmin@skku.edu,~yoonbai@skku.edu,~ddtolla@skku.edu} }

\end{center}
\vspace{15mm}

\begin{abstract}
The duality between a $d$-dimensional conformal field theory with relevant deformation and a gravity theory on an asymptotically AdS$_{d+1}$ geometry, has become a suitable tool in the investigation of the emergence of gravity from quantum entanglement in field theory. 
Recently, we have tested the duality between the mass-deformed ABJM theory and asymptotically AdS$_4$ gravity theory, which is obtained from the  KK reduction of the 11-dimensional supergravity on the LLM geometry. In this paper, we extend the KK reduction procedure beyond the linear order and establish non-trivial KK maps between 4-dimensional fields and 11-dimensional fluctuations.    
We rely on this gauge/gravity duality to calculate the entanglement entropy by using the Ryu-Takayanagi holographic formula  and the path integral method developed by Faulkner. We show that the entanglement entropies obtained using these two methods agree when the asymptotically AdS$_4$ metric satisfies the linearized Einstein equation with nonvanishing energy-momentum tensor for two scalar fields. These scalar fields encode the information of the relevant deformation of the ABJM theory. This confirms that the asymptotic limit of LLM geometry is the emergent gravity of the quantum entanglement in the mass-deformed ABJM theory with a small mass parameter. 
We also comment on the issue of the relative entropy and the Fisher information in our setup.
\end{abstract}

\newpage
\tableofcontents

\section{Introduction}

The holographic approach has become a very efficient technique in computing the entanglement entropy (EE) which is rather challenging in  a direct path integral approach in quantum field theories (QFTs) in more than two dimensions. 
The holographic calculation of the EE, which was  proposed by Ryu and Takayanagi (RT)~\cite{Ryu:2006bv,Ryu:2006ef} and its covariant generalization~\cite{Hubeny:2007xt}, drew much attention due to its elegance and implications in condensed matter physics and gravity theories.

More recently, the idea of the holographic entanglement entropy (HEE) caught more attention because of its importance in unlocking  some mysteries of the gauge/gravity correspondence~\cite{Maldacena:1997re,Gubser:1998bc} itself. 
The RT formula relates the EE ($S_A$) of a $(d-1)$-dimensional spatial  subregion $A$ in the vacuum state of a $d$-dimensional conformal field theory (CFT), which is living on the boundary of $(d + 1)$-dimensional AdS space, to a surface with  minimum area in the bulk of the AdS space with the same boundary as the subregion $A$. 
This creates a connection between the EE of QFT and the emergent spacetime geometry of the dual gravity theory. 
The topic of the emergence of gravity from quantum entanglement has shown  some significant progresses in recent years~\cite{Ryu:2006bv,Ryu:2006ef,Swingle:2009bg,VanRaamsdonk:2009ar}. 
Much of the progress in this direction esteemed from the first law of EE ($\delta S_A=\delta\langle H_A\rangle$)~\cite{Bhattacharya:2012mi,Allahbakhshi:2013rda,Blanco:2013joa,Wong:2013gua}, which equates the variation of EE due to a transition from  a vacuum state of a CFT  to some nearby state, and the variation of the vacuum expectation value ($vev$) of some characteristic Hamiltonian associated with the subregion $A$, which is known as the modular Hamiltonian. For a ball-shaped subregion $A$ on the boundary of an asymptotically AdS space, it was shown that the first law of EE is equivalent to a linearized Einstein equation on the AdS background~\cite{Lashkari:2013koa,Faulkner:2013ica,Swingle:2014uza}. Alternative approaches and extenstions of this phenomenon  were given in \cite{Lee:2010bg,Mosk:2016elb,Czech:2016tqr,Oh:2017pkr}. 
See also \cite{Faulkner:2017tkh} for the emergence of nonlinear gravitational equations from a broad class of CFTs via the EE.

In this paper, we pursue this phenomena of the emergent spacetime geometry in the context of the 3-dimensional mass-deformed Aharony-Bergman-Jefferis-Meldacena (mABJM) theory~\cite{Hosomichi:2008jb,Gomis:2008vc} and the dual gravity theory. 
The mABJM theory is obtained from the original ${\cal N}=6$ ABJM theory~\cite{Aharony:2008ug} in 3-dimensions by a relevant deformation which preserves the full supersymmetry. The dual gravity theory is the 11-dimensional supergravity on the Lin-Lunin-Maldacena (LLM) geometry~\cite{Lin:2004nb} with ${\mathbb Z}_k$ orbifold and SO(2,1)$\times$SO(4)/${\mathbb Z}_k\times$SO(4)/${\mathbb Z}_k$ isometry, which is asymptotically AdS$_4\times S^7/{\mathbb Z}_k$.
In \cite{Jang:2016tbk}, we have presented a compelling evidence for the gauge/gravity duality between the mABJM theory and the 11-dimensional gravity on the LLM geometry. 
This was achieved  by applying the gauge/gravity dictionary~\cite{Gubser:1998bc} to the $vev$s of the chiral primary operator (CPO) with conformal dimension $\Delta=1$ for all supersymmetric vacua of the mABJM theory and the 4-dimensional dual scalar modes obtained from the procedure of the Kaluza-Klein (KK) holography~\cite{Skenderis:2006uy,Skenderis:2006di,Skenderis:2007yb} of the 11-dimensional supergravity. We found an exact agreement between the results from the both sides in the large $N$ limit\footnote{See also \cite{Jang:2018aqr} for the extension to the $vevs$ of the CPO with conformal dimension $\Delta=2$.}. 

In order to show the exact dual relation in \cite{Jang:2016tbk}, we mainly dealt  with the matter fields, such as scalar, vector, and tensor fields, without considering the 4-dimensional metric. 
In this paper, however, we construct the 4-dimensional Einstein-Hilbert action with matter from the KK reduction of the 11-dimensional gravity on the LLM geometries with $\mathbb{Z}_k$ orbifold. The matter content of the 4-dimensional theory is determined by the asymptotic behaviour of the LLM solutions. At quadratic order in the mass parameter $\mu_0$ of the LLM solutions, the matter sector is composed of only one scalar field ($\Psi$) and one pseudoscalar field ($T$). These matter fields as well as the 4-dimensional graviton field $(H_{\mu\nu}$) are related to the 11-dimensional KK modes by some non-trivial field redefinitions, often called  the KK maps. Solution of the linearized Einstein equation for $(H_{\mu\nu})$ with the energy-momentum tensor for the two scalar fields is asymptotically AdS$_4$ and encodes the information of all LLM geometries in the small mass limit.

The presence of the two matter fields in the 4-dimensional gravity theory corresponds to the deformation of the ABJM theory by two relevant operators, which are a CPO of conformal dimension one (${\cal O}^{(1)}$) dual to $\Psi$ and a gauge invariant operator of conformal dimension two ($\tilde{\cal O}^{(2)}$) dual to $T$. 
In this setting, where the CFT with relevant deformation and its dual gravity theory  are explicitly known, we investigate the emergent gravity phenomena away from the UV fixed point by computing the variation of EE ($\delta S_A$) of the disk-shaped region $A$ in terms of the RT formula~\cite{Ryu:2006bv,Ryu:2006ef} in the gravity theory and the path integral method developed in \cite{Faulkner:2014jva} in QFT side.

The EE for a CFT with relevant deformation is calculated using the path integral method~\cite{Rosenhaus:2014woa,Rosenhaus:2014zza,Faulkner:2014jva,Faulkner:2015csl}, in which the EE is represented in terms of correlation functions. See also~\cite{Haehl:2015rza,Speranza:2016jwt,Beach:2016ocq,Sarosi:2017rsq}.
In particular, in \cite{Faulkner:2014jva} the author showed that the EE of the $d$-dimensional CFT with a relevant deformation can be regarded as a conserved charge in an emergent (auxiliary) $(d+1)$-dimensional gravity theory.  The same quantity was also computed in the dual gravity theory~\cite{Liu:2012eea,Nozaki:2013vta,Lin:2014hva,Lashkari:2015hha,Casini:2016rwj,Taylor:2016aoi} using the RT formula. 
The EE from the direct path integral approach is exactly the same as the HEE in RT formula with the metric satisfying the linearized Einstein equation in the presence of a nonvanishing energy-momentum tensor.

In our case, the deformations by the two relevant operators, ${\cal O}^{(1)}$ and $\tilde{\cal O}^{(2)}$, belong to the $\Delta< d/2$ and $\Delta>d/2$ cases, which require separate treatments \cite{Liu:2012eea,Faulkner:2014jva,Speranza:2016jwt}. 
In this setup, we calculate the variation of the EE ($\delta S_A$) of the disk region $A$ up to quadratic order in the deformation parameter in the QFT side and the gravity side separately. In the QFT side, we rely on the exact dual relation established in \cite{Jang:2016tbk} and use the method developed in \cite{Faulkner:2014jva}, while in the gravity side we use the RT formula for the 4-dimensional metric in the presence of the matter fields\footnote{The HEE obtained from the 4-dimensional metric agrees with the HEE which was obtained from the 11-dimensional LLM geometries before the KK reduction~\cite{Kim:2016dzw}. 
See also \cite{Kim:2014yca}. }.
We show that the $\delta S_A$'s obtained from both sides are equal only when the Einstein equation in the 4-dimensional gravity theory is satisfied. It implies that the asymptotic limit of the LLM geometry is actually the emergent gravity in the small mass expansion.

The remaining part of the paper is organized as follows. In section 2, we construct the 4-dimensional gravity from the KK reduction of the 11-dimensional supergravity. We also setup the non-trivial KK maps between the 11-dimensional fields and the 4-dimensional fields. In section 3, we use the RT formula to calculate the HEE from an asymptotically AdS$_4$ metric, which is obtained from the KK reduction of the LLM solution. We then establish the relation between the variation of the HEE and the $vev$ of a conformal dimension one CPO plus the $source$ of a conformal dimension two gauge invariant operator in mABJM theory. In section 4, we review the path integral methods necessary to obtain the EE in a CFT with some relevant deformation. We apply these methods to our setup and obtain the variation of the EE for mABJM theory at quadratic order in the deformation parameter. We use the results in the first law of EE at quadratic order, to show the emergence of the gravity in 4-dimensions in a top-down approach from the quantum entanglement of the 3-dimensional mABJM theory. In section 5, we draw some conclusions and discuss some future directions.

\section{Construction of 4-dimensional Gravity Theory}\label{4dgravity}
In this section, we construct a 4-dimensional gravity theory on AdS$_4$ background from the KK reduction of the 11-dimensional supergravity theory. In particular, we focus on a 4-dimensional gravity theory whose matter content is determined by the LLM solution with SO(2,1)$\times$SO(4)/${\mathbb Z}_k\times$SO(4)/${\mathbb Z}_k$ isometry.  In \cite{Jang:2016tbk} we performed the KK reduction at linear order and obtained the linearized KK maps among gauge invariant 11-dimensional and 4-dimensional fields. The linearized  reduction involves the truncation of the field equations at linear order in the fluctuations on the AdS$_4\times$S$^7$/${\mathbb Z}_k$ solution of the 11-dimensional supergravity.  The linearized field equations are solved by the asymptotic expansion of the LLM solutions at linear order in the mass parameter $\mu_0$. However, to solve the field equations at $\mu_0^2$ order or higher, the linearized KK reduction is not sufficient. In this section, we construct the nonlinear KK mapping up to $\mu_0^2$ order by truncating the field equations at quadratic order in the fluctuations.

\subsection{Field equations at quadratic order}
The functional variation of the bosonic part of the 11-dimensional supergravity action gives the following equations for the metric and the 3-form gauge field, 
\begin{align}\label{gmunuEoM}
&{\bf R}_{pq}-\frac12 {\bf g}_{pq} {\bf R}=\frac{1}{48}
\Big(-\frac12{\bf g}_{pq}{\bf F}_{rstu} {\bf F}^{rstu}+4 {\bf F}_{pstu} {\bf F}_q^{~stu}\Big),
\nn\\
&\partial_p(\sqrt{{\bf -g}}{\bf F}^{pqrs})+\frac{1}{2\cdot (4!)^2}\tilde\epsilon^{p_1\cdots p_4q_1\cdots q_4qrs}{\bf F}_{p_1\cdots p_4} {\bf F}_{q_1\cdots q_4}=0,
\end{align}
where we used the index notation
$(p,q,r,\cdots =0,1,\cdots,10)$ and ~$\tilde\epsilon^{012\cdots 10}=-1$ is the Levi-Civita symbol\footnote{We use a notation in which the quantities in 11-dimensional supergravity are denoted by bold font symbols whereas the normal font symbols are reserved for AdS$_4\times S^7$ values of those quantities.}.  The AdS$_4\times S^7$ solution of \eqref{gmunuEoM} is given by\footnote{For the ${\mathbb Z}_k$ orbifold, we discuss in the subsection \ref{Zk}.} 
\begin{align}\label{AdS4S7}
&ds^2=\frac{L^2}{ 4z^2}\left(-dt^2+dw_1^2+ dw_2^2+dz^2\right)+ L^2 ds^{2}_{S^{7}},\nn\\
&F_{\mu\nu\rho\sigma}=-\frac6{L}\epsilon_{\mu\nu\rho\sigma}, ~{\rm and~it~is~zero~otherwise},
\end{align}
where we split the 11-dimensional indices into the AdS$_4$ indices
$(\mu,\nu,\cdots =0,\cdots,3)$ and the $S^7$ indices $(a,b,\cdots=4,\cdots,10)$.  Here $\epsilon_{\mu\nu\rho\sigma} = \sqrt{|g_{{\rm AdS_4}}|}
\,\tilde \epsilon_{\mu\nu\rho\sigma}$ is the Levi-Civita tensor for the AdS$_4$ space and $L$ is the radius of $S^7$.

To obtain nonlinear field equations from \eqref{gmunuEoM}, we perturb the fields around AdS$_4\times S^7$ background by expressing the 11-dimensional metric and the 4-form field strength as 
\begin{align}\label{fluct}
{\bf g}_{pq}=g_{pq}+h_{pq}, \qquad {\bf F}_{pqrs}=F_{pqrs}+f_{pqrs},
\end{align}
and keep all terms up to quadratic order in the fluctuations $h_{pq}$ and $f_{pqrs}$. Applying such perturbation to the metric equation in \eqref{gmunuEoM}, we obtain
\begin{align}
&\nabla^r\nabla_{p}h_{qr}+\nabla^r\nabla_{q}h_{pr}-\nabla^2h_{pq}-\nabla_q\nabla_ph^r{}_{r}-Rh_{pq}-g_{pq}\left(-R^{rs}h_{rs}
+\nabla^r\nabla^sh_{rs}-\nabla^2h^r{}_{r}\right)
\nn\\
&+\frac{1}{48}\Big(F_{rstu}F^{rstu} h_{pq}
{-4}g_{pq}h_{rs}F^r{}_{tuv}F^{stuv}\Big)+\frac{1}{24}g_{pq}
f_{pqrs}F^{pqrs}-\frac12 h_{rs}F^r{}_{ptu}F_q{}^{stu}\nn\\
&-\frac1{6}\Big(
f_{prst}F_q^{~rst}+F_{prst}f_q^{~rst}\Big)+Q_{pq}=0,
\label{lineq2-1}
\end{align}
where the indices are raised (lowered) by the AdS$_4\times S^7$ metric, and the covariant derivatives are also those of the background. Here $Q_{pq}$ denotes terms which are quadratic in the fluctuations,
\begin{align}\label{lineq1-2}
Q_{pq}=&
-\nabla_r\Big(h^{rs}\big[\nabla_p h_{sq}
+\nabla_q h_{sp}-\nabla_s h_{pq}\big]\Big)+\frac12\nabla_qh^{rs}\nabla_p h_{rs}+h^{rs}\nabla_q\nabla_p h_{rs}\nn\\
&+\frac12\nabla^rh^s{}_s\big[\nabla_p h_{rq}+\nabla_q h_{rp}-\nabla_r h_{pq}\big]
+\nabla^r h^s{}_{q}\nabla_r h_{sp}-\nabla^r h^s{}_{q}\nabla_s h_{pr}-g_{pq}R_{rs}h^{rt}h^s{}_t\nn\\
&+\frac12g_{pq}\nabla_r\Big(h^{rs}\big[2\nabla^t h_{st}-\nabla_s h^t{}_t\big]\Big)-\frac34g_{pq}\nabla^th^{rs}\nabla_t h_{rs}+\frac12g_{pq}\nabla^r h^{st}\nabla_s h_{tr}-\frac12g_{pq}h^{rs}\nabla^2 h_{rs}\nn\\ 
&-\frac14g_{pq}\nabla^rh^s{}_s\big[2\nabla^s h_{rs}-\nabla_r h^t{}_t\big]
+\frac12g_{pq}h^{rs}\Big(\nabla^t\nabla_rh_{ts}+\nabla^t\nabla_sh_{tr}-\nabla^2h_{rs}-\nabla_r\nabla_sh^t{}_t\Big)\nn\\
&+h_{pq}h^{rs}R_{rs}-h_{pq}\Big(\nabla^r\nabla^sh_{rs}-\nabla^2h^r{}_r\Big)\\
+\frac1{12}&\Big(g_{pq}F_{rstu}F^{rst}~\!\!_{s'}h^{uv}h_v^{s'}+\frac32g_{pq}F_{rstu}F^{rs}~\!\!_{r's'}h^{tr'}h^{us'}-g_{pq}F_{rstu}f^{rst}~\!\!_{s'}h^{us'}\nn\\
&+\frac12h_{pq}f_{rstu}F^{rstu}-h_{pq}F_{rstu}F^{rst}~\!\!_{s'}h^{us'}+\frac14g_{pq}f_{rstu}f^{rstu}-g_{pq}f_{rstu}F^{rst}~\!\!_{s'}h^{us'}\Big)\nn\\
-\frac1{2}&\Big(F_{pstu}F_q^{~st}~\!\!_{s'}h^{uv}h_v{}^{s'}+F_{pstu}F_q^{~s}~\!\!_{r's'}h^{tr'}h^{us'}-F_{pstu}f_q^{~st}~\!\!_{s'}h^{us'}-f_{pstu}F_q^{~st}~\!\!_{s'}h^{us'}+\frac13f_{pstu}f_q^{~stu}\Big).\nn
\end{align}
Similarly, from the gauge field equation in \eqref{gmunuEoM} we obtain
\begin{align}\label{C3EoM-1}
\nabla_p(h^{t}{}_{t}F^{pqrs})+2\nabla_p(4F_{s'}^{~[pqr}h^{s]s'}+f^{pqrs})
&+\frac2{\sqrt{-g}}\frac{1}{(4!)^2}\tilde\epsilon^{p_1\cdots p_4q_1\cdots q_4qrs}f_{p_1\cdots p_4}F_{q_1\cdots q_4}+P^{qrs}=0,
\end{align} 
where 
\begin{align}\label{C3EoM-2}
P^{qrs}=&-\frac12\nabla_p\Big[\Big(h_{tu}h^{tu}-\frac12(h^{t}{}_{t})^2\Big) F^{pqrs}\Big]-8\nabla_p\Big[F_{s'}^{~[pqr}h^{s]t}h_t^{s'}-\frac32F^{r's'[pq}~\!\!h^r{}_{r'}h^{s]}{}_{s'}-f_{s'}^{~[pqr}h^{s]s'}\Big]\nn\\
&+\nabla_p\Big[h^{t}{}_{t}\big( 4F_{s'}^{~[pqr}h^{s]s'}+f^{pqrs}\big)\Big]+\frac1{\sqrt{-g}}\frac{1}{(4!)^2}\tilde\epsilon^{p_1\cdots p_4q_1\cdots q_4qrs}f_{p_1\cdots p_4}f_{q_1\cdots q_4}.
\end{align} 

For later convenience we write \eqref{lineq2-1} and \eqref{C3EoM-1} by separating the AdS$_4$  and  the $S^7$ indices and also inserting the AdS$_4\times S^7$ background information given in \eqref{AdS4S7}. One can write those quadratic equations in generic gauge, however, having the LLM solutions in mind, we simplify the equations by setting $h_{\mu a}$ and $f_{\mu\nu ab}$ to zero. Then we obtain
\begin{align}\label{lineq2-5}
&\nabla^\rho\nabla_{\mu}h_{\nu\rho}+\nabla^\rho\nabla_{\nu}h_{\mu\rho}
-(\nabla^\rho\nabla_\rho+\nabla^a\nabla_a) h_{\mu\nu}-\nabla_\mu\nabla_\nu (h^\rho_{~\rho}+h^a_{~a})+\frac{24}{L^2}h_{\mu\nu}\nn\\
&-g_{\mu\nu}\left[\frac{30}{L^2}h^\rho_{~\rho}-\frac6{L^2}h^a_{~a}+\nabla^\rho\nabla^\sigma h_{\rho\sigma}+\nabla^a\nabla^b h_{ab}-(\nabla^\rho\nabla_\rho+\nabla^a\nabla_a) (h^\sigma_{~\sigma}+h^b_{~b})\right]\nn\\
&-\frac1{4L}g_{\mu\nu}f_{\rho\sigma\tau\lambda}\epsilon^{\rho\sigma\tau\lambda}+\frac1{L}\Big(f_{\mu \rho\sigma\tau}\epsilon_\nu^{~\rho\sigma\tau}+f_{\nu \rho\sigma\tau}\epsilon_\mu^{~\rho\sigma\tau}\Big)+Q_{\mu\nu}=0,
\end{align}
\begin{align}\label{lineq2-6}
&\nabla^\rho\nabla_{a}h_{\mu\rho}
+\nabla^b\nabla_{\mu}h_{a b}-\nabla_\mu\nabla_a (h^\rho_{~\rho}+h^b_{~b})+\frac1{L}f_{a \rho\sigma\tau}\epsilon_\mu^{~\rho\sigma\tau}+Q_{\mu a}=0,
\end{align}
\begin{align}\label{lineq2-7}
&\nabla^c\nabla_{a}h_{b c}+\nabla^c\nabla_{b}h_{a c}-(\nabla^\rho\nabla_\rho+\nabla^c\nabla_c) h_{ab}-\nabla_a\nabla_b (h^\rho_{~\rho}+h^c_{~c})-\frac{12}{L^2}h_{ab}-\frac1{4L}g_{ab}f_{\rho\sigma\tau\lambda}\epsilon^{\rho\sigma\tau\lambda}\nn\\
&-g_{ab}\left(-\frac{6}{L^2}h^\rho_{~\rho}-\frac6{L^2}h^c_{~c}+\nabla^\rho\nabla^\sigma h_{\rho\sigma}+\nabla^c\nabla^d h_{cd}-(\nabla^\rho\nabla_\rho+\nabla^c\nabla_c) (h^\sigma_{~\sigma}+h^d_{~d})\right)+Q_{ab}=0,
\end{align}
and 
\begin{align}
&\nabla_\sigma f^{\sigma\mu\nu\rho}+\nabla_a f^{a\mu\nu\rho}
-\frac3L(\nabla_\sigma h^\lambda_{~\lambda})\epsilon^{\sigma\mu\nu\rho}-\frac3L(\nabla_\sigma h^a_{~a})F^{\sigma\mu\nu\rho}-\frac{24}L\nabla_\lambda\big(h_\sigma^{[\lambda}\epsilon^{\mu\nu\rho]\sigma}\big)+P^{\mu\nu\rho}=0,\nn\\
&\nabla_\sigma f^{\sigma\mu\nu a}+P^{\mu\nu a}=0\nn,\qquad \nabla_c f^{c\mu ab}+P^{\mu ab}
=0,\nn\\
&\nabla_\sigma f^{\sigma abc}+\nabla_d f^{dabc}
+\frac{1}{(4!)^2}\epsilon^{a_1\cdots a_4\nu_1\cdots \nu_4 abc}f_{a_1\cdots a_4}F_{\nu_1\cdots \nu_4}+P^{abc}=0.\label{C3EoM4-2}
\end{align} 
\subsection{Expansion in spherical harmonics}\label{Zk}

The KK reduction involves the expansions of the fluctuations $h_{pq}$ and $f_{pqrs}$ in terms of the spherical harmonics on $S^7$. Here we are interested in the asymptotic limit of the LLM geometry with SO(2,1)$\times {\rm SO}(4)/{\mathbb Z}_k \times {\rm SO}(4)/{\mathbb Z}_k$ isometry. In that case, we need to consider expansion in terms of  the spherical harmonics on $S^7/\mathbb{Z}_{k}$ with $ {\rm SO}(4)/{\mathbb Z}_k \times {\rm SO}(4)/{\mathbb Z}_k$ symmetry.  In the presence of such symmetry, the  metric on the $S^7/\mathbb{Z}_{k}$ is written as
\begin{align}
ds^2_{S^7/\mathbb{Z}_{k}}=d\tau^2+\frac{d\theta^2 +\sin^2\theta d\phi^2+(d\psi 
+\cos\theta d\phi)^2}4+\frac{d\tilde\theta^2 +\sin^2\tilde\theta d\tilde\phi^2+(d\tilde\psi +\cos\tilde\theta d\tilde\phi)^2}4,\nn
\end{align}
with ranges of the angles, $ 0\le \theta,\tilde\theta \le \pi, \,\, 0\le \phi,\tilde\phi \le 2\pi,\,\, 0\le\psi,\tilde\psi \le \frac{4\pi}{k}$.   The $\mathbb{Z}_k$ orbifolding acts as $\left(\psi, \, \tilde\psi\right) \to \left( \psi + \frac{4\pi}{k},\, \tilde\psi + \frac{4\pi}{k}\right)$\cite{Auzzi:2009es,Cheon:2011gv}.  The spherical harmonics with the $ {\rm SO}(4)/{\mathbb Z}_k \times {\rm SO}(4)/{\mathbb Z}_k$ symmetry are dependent only on the $\tau$ coordinate, and they are the same with and without the orbifolding. This implies that expansions of the fluctuations $h_{pq}$ and $f_{pqrs}$ in terms of these spherical harmonics are the same with and without the orbifolding. 
 
In \cite{Jang:2016tbk}, we have written a complete form of the expansions in terms of the spherical harmonics on $S^7$. For clarity of presentation, let us recall the expansion,
\begin{align}\label{hpqexp}
&
h_{\mu\nu}(x,y)
=
h^{I_1}_{\mu\nu}(x)Y^{I_1}(y),
\nn\\
&h_{\mu a}(x,y)
=
v_\mu^{I_7}(x)Y_a^{I_7}(y)
+s^{I_1}_\mu(x)\nabla_{a}Y^{I_1}(y),
\nn\\
&
h_{(ab)}(x,y)
=
t^{I_{27}}(x)Y_{(ab)}^{I_{27}}(y)
+v^{I_7}(x)\nabla_{(a}Y_{b)}^{I_7}(y)
+s^{I_1}(x)\nabla_{(a}\nabla_{b)}Y^{I_1}(y),
\nn\\
&
h^a_{~a}(x,y)
=
\phi^{I_1}(x)Y^{I_1}(y), 
\nn \\
&
f_{\mu\nu\rho\sigma}(x,y)
=
4\nabla_{[\mu}s_{\nu\rho\sigma]}^{I_1}(x)Y^{I_1}(y),
\nn\\
&
f_{\mu\nu\rho a}(x,y)
=
3\nabla_{[\mu}v_{\nu\rho]}^{I_7}(x)Y_a^{I_7}(y)
-s^{I_1}_{\mu\nu\rho}(x)\nabla_{a}Y^{I_1}(y),
\nn\\
&
f_{\mu\nu ab}(x,y)
=
2\nabla_{[\mu}t_{\nu]}^{I_{21}}(x)Y_{[ab]}^{I_{21}}(y)
+2v_{\mu\nu}^{I_7}(x)\nabla_{[a}Y_{b]}^{I_7}(y)
,\nn\\
&
f_{\mu abc}(x,y)
=
\nabla_\mu t^{I_{35}}(x)Y_{[abc]}^{I_{35}}(y)
-3 t_\mu^{I_{21}}(x)\nabla_{[a}Y_{bc]}^{I_{21}}(y)
,\nn\\
&
f_{abcd}(x,y)
=
4 t^{I_{35}}(x)\nabla_{[a}Y_{bcd]}^{I_{35}}(y),
\end{align}
where $x$ denotes the AdS${}_4$ coordinates and $y$ denotes the $S^7$ coordinates. For the details about the spherical harmonics on $S^7$, see \cite{Jang:2016tbk}. The parenthesis of two indices $(ab)$ means symmetrized traceless combination, while  the square bracket $[ab\cdots]$ denotes complete antisymmetrization of indices. Eventually, we will identify the fluctuations $h_{pq}$ and $f_{pqrs}$ with the deviations of the LLM geometries from the AdS$_4\times S^7$ background. In the gauge choice of the LLM solutions, $h_{\mu a}$ and $f_{\mu\nu ab}$ are zero and as a result most of the KK towers  in \eqref{hpqexp} are absent.  Therefore, we use a truncated  expansion,
\begin{align}\label{metric-exp1}
&h_{\mu\nu}(x,y)=h^{I_1}_{\mu\nu}(x)Y^{I_1}(y),\quad h^\rho{}_{\rho}(x,y)=h^{I_1}(x)Y^{I_1}(y),\nn\\
&h^a_{~a}(x,y)=\phi^{I_1}(x)Y^{I_1}(y),\quad h_{(ab)}=s^{I_1}(x)\nabla_{(a}\nabla_{b)}Y^{I_1}(y),
\end{align}
and
\begin{align}\label{F4-exp1}
&f_{\mu\nu\rho\sigma}(x,y)=4\nabla_{[\mu} s_{\nu\rho\sigma]}^{I_1}(x)Y^{I_1}(y),\quad f_{\mu\nu\rho a}(x,y)=- s^{I_1}_{\mu\nu\rho}(x)\nabla_{a}Y^{I_1}(y),\nn\\
&f_{\mu abc}(x,y)=\nabla_\mu  t^{I_{35}}(x)Y_{abc}^{I_{35}}(y)
,\quad f_{abcd}(x,y)=4 t^{I_{35}}(x)\nabla_{[a}Y_{bcd]}^{I_{35}}(y).
\end{align}
This truncation is not dictated by some symmetry, which would have been a requirement in order to have a consistent truncation if one follows the line of thought of ref.~\cite{Cvetic:2000dm}. However, the equations obtained from the truncated expansion are consistent at quadratic order in the fluctuations. The reason is that those equations are solved order by order in the mass parameter ($\mu_0$) of the LLM solutions, and the modes which are omitted from the expansion are all vanishing, at least up to $\mu_0^2$-order~\cite{Jang:2016tbk}.

Plugging \eqref{metric-exp1} and \eqref{F4-exp1} into \eqref{lineq2-5} and then projecting on the scalar harmonics $Y^{I_1}$, we obtain
\begin{align}\label{lineqmn}
&-\left(\square+\Lambda^{I_1}-\frac{24}{L^2}\right) h^{I_1}_{\mu\nu}+\nabla^\rho\nabla_{\mu}h^{I_1}_{\nu\rho}+\nabla^\rho\nabla_{\nu}h^{I_1}_{\mu\rho}
-\nabla_\mu\nabla_\nu (h^{I_1}+\phi^{I_1})\nn\\
&+g_{\mu\nu}\left(\square+\Lambda^{I_1}-\frac{30}{L^2}\right)h^{I_1}
+g_{\mu\nu}\left(\square+\frac67\Lambda^{I_1}+\frac6{L^2}\right)\phi^{I_1}-g_{\mu\nu}\left(\frac67\Lambda^{I_1}+\frac6{L^2}\right)\Lambda^{I_1}s^{I_1}\nn\\
&-g_{\mu\nu}\nabla^\rho\nabla^\sigma h^{I_1}_{\rho\sigma}+\frac{1}{L}g_{\mu\nu}\nabla^{\rho} t^{I_1}_\rho+Q^{I_1}_{\mu\nu}=0,
\end{align}
 where $\square\equiv\nabla_\mu\nabla^\mu$, ~$Q^{I_1}_{\mu\nu}=\frac1{\omega_7}\int_{S^7}Q_{\mu\nu}Y^{I_1}$ with the unit volume of the $S^7$, $\omega_7$, and we have set $s^{I_1}_{\mu\nu\rho}=\frac1{3!}\epsilon_{\mu\nu\rho}{}^{\sigma}t^{I_{1}}_\sigma$, and $\Lambda^{I_{1}}=-\frac{I_{1}(I_{1}+6)}{L^{2}}$ is the eigenvalue corresponding to the scalar harmonics $Y^{I_1}$. Taking the trace of the above equation, we obtain
 \begin{align}\label{h-eqn}
&\Big(2\square+3\Lambda^{I_1}-\frac{96}{L^2}\Big)h^{I_1}-2\nabla^\mu\nabla^\nu h^{I_1}_{\mu\nu}+3\Big(\square+\frac87\Lambda^{I_1}+\frac8{L^2}\Big)\phi^{I_1}+\frac4{L}\nabla^{\rho} t^{I_1}_\rho\nn\\
&-24\Lambda^{I_1}\Big(\frac17\Lambda^{I_1}
+\frac1{L^2}\Big)s^{I_1}+Q^{I_1}=0
\end{align}
with $Q^{I_1}=g^{\mu\nu}Q_{\mu\nu}^{I_1}$. 
From \eqref{lineq2-6}, we obtain  the following equation by projecting on $\nabla^a Y^{I_1}$  with $I_1\ne0$,
\begin{align}\label{leq2-2-2}
&-\Big(\frac67 \Lambda^{I_1}+ \frac6{L^2}\Big)\nabla_\mu s^{I_1}+\frac67\nabla_\mu\phi^{I_1}
-\nabla^\nu h_{\mu\nu}^{I_1}+\nabla_\mu h^{I_1}
-\frac1{L}  t^{I_1}_\mu+Q_{\mu}^{I_1}
=0,
\end{align}
where $Q_{\mu}^{I_1}=\frac1{\omega_7}\int_{S^7}Q_{\mu a}\nabla^aY^{I_1}$.
From \eqref{lineq2-7}, we obtain two scalar equations by projecting on  $ g^{ab} Y^{I_1}$ and $\nabla^{(a}\nabla^{b)} Y^{I_1}$, in the latter case $I_1\ne 0$,
\begin{align}
&
\frac67\Big(\square+\frac{5}7\Lambda^{I_1}+\frac{5}{L^2}\Big)\phi^{I_1}
+\Big(\square+\frac67\Lambda^{I_1}+\frac{6}{L^2}\Big)h^{I_1}
-\nabla^\mu\nabla^\nu h_{\mu\nu}^{I_1}
-\frac1{L}\nabla^{\rho} t^{I_1}_\rho\nn\\
&
-\frac{30}7\Lambda^{I_1}\Big(\frac{\Lambda^{I_1}}7+\frac1{L^2}
\Big)s^{I_1}+\tilde Q^{I_1}=0,\label{leq3-4-2}\\
&6\Lambda^{I_1}\Big(\frac{\Lambda^{I_1}}7+\frac1{L^2}
\Big)\Big[\Big(\square-\frac57\Lambda^{I_1}\Big)s^{I_1}+h^{I_1}+\frac57\phi^{I_1}\Big]-\hat Q^{I_1}=0
,\label{leq3-3-2}
\end{align}
where $\tilde Q^{I_1}=\frac1{\omega_7}\int_{S^7}Q_{ab}g^{ab}Y^{I_1}$ and $\hat Q^{I_1}=\frac1{\omega_7}\int_{S^7}Q_{ab}\nabla^{(a}\nabla^{b)} Y^{I_1}$. 

Similarly, inserting \eqref{metric-exp1} and \eqref{F4-exp1} into \eqref{C3EoM4-2},  and projecting on the appropriate spherical harmonic elements, we obtain the following set of equations
\begin{align}
&4\nabla^{\sigma}\nabla_{[\sigma} s^{I_1}_{\mu\nu\rho]}+\Lambda^{I_1}s^{I_1}_{\mu\nu\rho}-\frac3{L}\epsilon_{\sigma\mu\nu\rho}\nabla^\sigma\big( h^{I_1}+\phi^{I_1}\big)-\frac{24}{L}\nabla^\sigma h^{I_1}_{\lambda[\sigma}\epsilon_{\mu\nu\rho]}~\!\!^{\lambda}+P_{\mu\nu\rho}^{I_1}=0,\label{tSmnr1}\\
&\Lambda^{I_1}\nabla^{\rho} s^{I_1}_{\rho\mu\nu}+P_{\mu\nu}^{I_1}=0,\quad(I_1\ne0),\label{tSmn1} \\
&\left(\square-\frac{12}{L^2}+\Lambda^{I_{35}}\pm\frac{6(I_{35}+3)}{L^2}\right)t_{\pm}^{I_{35}}+P^{I_{35}}=0, \label{tT35}
\end{align}
where $\Lambda^{I_{35}}=-\frac{I_{35}(I_{35}+6)-3}{L^{2}}$ and
\begin{align}
&P_{\mu\nu\rho}^{I_1}=\frac1{\omega_7}\int_{S^7}P_{\mu\nu\rho}Y^{I_1},\qquad P_{\mu\nu}^{I_1}=\frac1{\omega_7}\int_{S^7}P_{\mu\nu a}\nabla^aY^{I_1},\qquad P^{I_{35}}=\frac1{\omega_7}\int_{S^7}P^{abc}Y_{abc}^{I_{35}}.
\end{align}
Here we have used the relation,
$\epsilon_{abc}~\!\!^{a_1a_2a_3a_4}\nabla_{a_1}Y^{I_{35}}_{a_2a_3a_4}
=\pm3!\frac{(I_{35}+3)}{L}Y^{I_{35}}_{abc}$, to obtain the  two equations in \eqref{tT35}.
For later convenience let us again set  $s^{I_1}_{\mu\nu\rho}=\frac1{3!}\epsilon_{\mu\nu\rho}^{~~~~\!\!\lambda}~\!t^{I_1}_\lambda$ and then multiply \eqref{tSmnr1} by $\epsilon_{\mu'}^{\!~~\mu\nu\rho}\nabla_{\nu'}$.  Thus we obtain 
\begin{align}\label{C3EoM-4b}
&\frac{18}L\nabla_{\mu}\nabla_{\nu}(-h^{I_1}+\phi^{I_1})+\Lambda^{I_1}\nabla_\nu t^{I_1}_\mu+\nabla_{\mu}\nabla_{\nu}\nabla^\rho t^{I_1}_\rho+\tilde P^{I_1}_{\mu\nu}=0,
\end{align}
where $\tilde P^{I_1}_{\mu\nu}=\epsilon_{\mu}^{\!~~\rho\sigma\lambda}\nabla_{\nu}P^{I_1}_{\rho\sigma\lambda}$. The trace of the above equation gives 
\begin{align}\label{C3EoM-4c}
&\frac{18}L\square(-h^{I_1}+\phi^{I_1})+(\square+\Lambda^{I_1})\nabla^\rho t^{I_1}_\rho+\tilde P^{I_1}=0,\qquad \tilde P^{I_1}=g^{\mu\nu}\tilde P^{I_1}_{\mu\nu}.\end{align}

\subsection{The 4-dimensional graviton equation at $\mu_0^2$ order}
From the quadratic equations in the previous subsection, one can obtain the quadratic order equations of motions for various 4-dimensional gauge invariant KK modes. In \cite{Jang:2016tbk} we have obtained the complete 4-dimensional KK spectrum, which in general is composed of three towers of scalar modes, two towers of  pseudoscalar modes, two towers of  vector modes, one tower of  pseudovector mode, and one tower of spin-two mode. We have also found the linear order equations for these modes in generic gauge. 
In this paper, we have set $h_{\mu a}$ and $f_{\mu\nu ab}$ to zero and as a result some of the KK towers disappear.
Furthermore, to construct the equation of motion for 4-dimensional graviton, we focus on the KK zero modes. To that end, we start from the zero modes of the equations \eqref{lineqmn}, \eqref{h-eqn}, \eqref{leq3-4-2}, \eqref{C3EoM-4b}, and \eqref{C3EoM-4c}. Let us rearrange those equations by introducing $u^{I_1}\equiv L\nabla^\rho t^{I_1}_\rho$ and $\hat\psi^{I_1}\equiv 18 h^{I_1}-u^{I_1}$, as follows
\begin{align}\label{lineqmn1}
&
\square h^0_{\mu\nu}=\frac{24}{L^2}h^0_{\mu\nu}+\nabla^\rho\nabla_{\mu}h^0_{\nu\rho}+\nabla^\rho\nabla_{\nu}h^0_{\mu\rho}
-\nabla_\mu\nabla_\nu (h^0+\phi^0)-g_{\mu\nu}\frac{4}{3L^2}\hat\psi^0\nn\\
&
\qquad\quad+Q^{0}_{\mu\nu}-\frac19 g_{\mu\nu}(Q^0+7\tilde Q^0),\\
&
\square h^0=-\frac{78}{L^2}h^0+\frac6{L^2}\phi^0+\frac5{L^2}u^0+\nabla^\rho\nabla^\sigma h^0_{\rho\sigma}+\frac13(2Q^{0}-7\tilde Q^0),\label{leq3-4-2a}\\
&
\square\phi^{0}=\frac{14}{3L^2}\hat\psi^0-\frac{12}{L^2}\phi^0-\frac79(Q^{0}-2\tilde Q^0),\label{h-eqn1}\\
& 
\square\hat\psi^0=\frac{84}{L^2}\hat\psi^0-\frac{216}{L^2}\phi^0-14(Q^{0}-2\tilde Q^0)+L\tilde P^{0},\label{C3EoM-4d}\\
&
-\nabla_{\mu}\nabla_{\nu}\hat\psi^0+18\nabla_{\mu}\nabla_{\nu}\phi^0+L\tilde P^{0}_{\mu\nu}=0.\label{C3EoM-4e}\end{align}

We notice that, the linear part of \eqref{lineqmn1} is not the 4-dimension linearized Einstein equation, which is given by
\begin{align}
\left(L_{E}+\frac{12}{L^{2}}\right)h^{0}_{\mu\nu}=0
\end{align}
with $L_{E}h^{0}_{\mu\nu}=\frac{1}{2}\left(-\square h^0_{\mu\nu}+\nabla^\rho\nabla_{\mu}h^0_{\nu\rho}+\nabla^\rho\nabla_{\nu}h^0_{\mu\rho}-\nabla_\mu\nabla_\nu h^0\right)$, where $L_{E}$ is the Einstein operator. Therefore, $h^0_{\mu\nu}$ is not the correct 4-dimensional graviton field. Neglecting the quadratic terms in equations \eqref{lineqmn1}-\eqref{C3EoM-4e}, the combination which satisfies the 4-dimensional linearized Einstein equation is  
\begin{align}
\hat h^0_{\mu\nu}\equiv h^0_{\mu\nu}-\frac14g_{\mu\nu}\phi^0+\frac1{24}g_{\mu\nu}\hat\psi^0. 
\end{align}
However, when we take into account the quadratic terms in \eqref{lineqmn1}-\eqref{C3EoM-4e}, $\hat h^0_{\mu\nu}$ still does not represent the correct 4-dimensional graviton field. In order to obtain the 4-dimensional graviton field, we need non-trivial field redefinitions to absorb the quadratic terms in the above equations of motion. Such field redefinition will be presented in the next subsection.   Now we combine  equations \eqref{lineqmn1}-\eqref{C3EoM-4e} to obtain the following quadratic equation for  $\hat h^0_{\mu\nu}$: 
\begin{align}\label{lineq1-3-03}
\square \hat h^0_{\mu\nu}=&\nabla^\rho\nabla_\mu h^0_{\rho\nu}+\nabla^\rho\nabla_\nu h^0_{\rho\mu}-\nabla_\mu\nabla_\nu h^0+\frac12\nabla_\mu\nabla_\nu\phi^0-\frac1{12}\nabla_{\mu}\nabla_{\nu}\hat\psi^0+\frac{24}{L^2}\hat h^0_{\mu\nu}\nn\\
&+Q^0_{\mu\nu}-\frac{L}{12}\tilde P^0_{\mu\nu}-\frac12g_{\mu\nu}\left(Q^0-\frac L{12}\tilde P^0\right).
\end{align}
Using the Einstein operator, we write \eqref{lineq1-3-03} as 
\begin{align}\label{lineq1-3-03a}
\left(L_E+\frac{12}{L^2}\right)\hat h^0_{\mu\nu}+\frac12Q^0_{\mu\nu}-\frac{L}{24}\tilde P^0_{\mu\nu}-\frac14g_{\mu\nu}\left(Q^0-\frac L{12}\tilde P^0\right)=0.
\end{align}

The explicit forms  of the quadratic terms ($Q^0_{\mu\nu}, \tilde P^0_{\mu\nu}$ etc.) are too long to display here, however, we would like to note that they all contain the terms which are quadratic in the fields $h^{I_1}_{\mu\nu}, h^{I_1},\phi^{I_1}, u^{I_1}, s^{I_1},t_\mu^{I_1}, t^{I_{35}}$ and their derivatives. In \cite{Jang:2016tbk} we have obtained the asymptotic expansion of the values of these fields in the LLM solutions. We have shown that, except for the modes with ${I_1}=2$ and $I_{35}=1$, the asymptotic expansions of the other modes, including the zero modes, are nonlinear in the expansion parameter $\mu_0$. In particular, the leading terms of all the zero modes that appear in the above equations are quadratic in $\mu_0$. Therefore, in order to solve the equations of motion of those zero modes at $\mu_0^2$ order, the quadratic terms in the above equations are built only by the modes with ${I_1}=2$ and $I_{35}=1$. Having said that, we can simplify \eqref{lineq1-3-03a} by using the linearized equations of motion for the ${I_1}=2$ and $I_{35}=1$ modes, which can be read from the list of equations in the previous subsection. 
\begin{align}
&\square h^2_{\mu\nu}=\frac{40}{L^2}h^2_{\mu\nu}+\nabla^\rho\nabla_\mu h^2_{\rho\nu}+\nabla^\rho\nabla_\nu h^2_{\rho\mu}-\nabla_\mu\nabla_\nu h^2-\nabla_\mu\nabla_\nu \phi^2-\frac4{3L^2}g_{\mu\nu}\hat\psi^2,\nn\\
&\square\phi^2=\frac{100}{L^2}h^2+\frac{108}{7L^2}\phi^2+\frac{1728}{7L^4}s^2-\frac{14}{3L^2}u^2,\quad \square \hat\psi^2=\frac{100}{L^2}\hat\psi^2+\frac{1944}{7L^2}\phi^2+\frac{31104}{7L^4}s^2,\nn\\
&\square s^2=-\frac{80}{7L^2}s^2-\frac57\phi^2-h^2,\quad t^2_{\mu}=\frac{54}{7L}\nabla_\mu s^2+\frac{6L}7\nabla_\mu\phi^2-L\nabla^\nu h^2_{\mu\nu}+L\nabla_\mu h^2,\nn\\
&\square t^1_{+}=-\frac8{L^2} t^1_{+}.
\end{align}
Here we picked only $t^1_+$ from the $t^{I_{35}=1}_\pm$ pair because the leading term in the asymptotic expansion of $t^1_-$ is cubic in $\mu_0$. Then we obtain
\begingroup
\allowdisplaybreaks
\begin{align}\label{lineq1-3-06}
&\Big(L_E+\frac{12}{L^2} \Big)\hat h_{\mu\nu}^0+\frac1{40}\Big\{{\frac3{8}}\big(\nabla_{\mu}\nabla_{\nu}h^2h^2-\nabla_{\mu}\nabla_{\nu}h^2\phi^2-h^2\nabla_{\mu}\nabla_{\nu}\phi^2+\frac{19}{21}\nabla_{\mu}\nabla_{\nu}\phi^2\phi^2\big)\nn\\
&+\frac1{48}\Big(-\nabla_{\mu}\nabla_{\nu}h^2u^2-h^2\nabla_{\mu}\nabla_{\nu}u^2\nn-\nabla_{\mu}h^2\nabla_{\nu}u^2-\nabla_{\nu}h^2\nabla_{\mu}u^2+\nabla_{\mu}\nabla_{\nu}\phi^2u^2\nn\\
&+\phi^2\nabla_{\mu}\nabla_{\nu}u^2+\nabla_{\mu}\phi^2\nabla_{\nu}u^2+\nabla_{\nu}\phi^2\nabla_{\mu}u^2\Big)+\frac{11}{72}\nabla_\mu h^2\nabla_\nu h^2+\frac{55}{392}\nabla_\mu \phi^2\nabla_\nu \phi^2\nn\\
&-\frac{3672}{49L^4}\nabla_\mu s^2\nabla_\nu s^2-\frac {95}{168}\big(\nabla_\mu h^2\nabla_\nu\phi^2+\nabla_\nu h^2\nabla_\mu\phi^2\big)-\frac {12}{7L^2}\big(\nabla_\mu h^2\nabla_\nu s^2+\nabla_\nu h^2\nabla_\mu s^2\big)\nn\\
&-\frac {72}{49L^2}\big(\nabla_\mu \phi^2\nabla_\nu s^2+\nabla_\nu \phi^2\nabla_\mu s^2\big)-\frac29\nabla^\rho h^2_{\mu\rho}\nabla^\sigma h^2_{\sigma\nu}+\frac29\big(\nabla_\mu h^2\nabla^\rho h^2_{\rho\nu}+\nabla_\nu h^2\nabla^\rho h^2_{\mu\rho}\big)\nn\\
&+\frac4{21}\big(\nabla_\mu \phi^2\nabla^\rho h^2_{\rho\nu}+\nabla_\nu \phi^2\nabla^\rho h^2_{\mu\rho}\big)+\frac{12}{7L^2}\big(\nabla_\mu s^2\nabla^\rho h^2_{\rho\nu}+\nabla_\nu s^2\nabla^\rho h^2_{\mu\rho}\big)+\nabla_\mu h^{2\rho\sigma}\nabla_\nu h^2_{\rho\sigma}\nn\\
&-\frac{1}2\nabla_\rho\big( h^{2\rho\sigma}\nabla_\mu h^2_{\sigma\nu}
+ h^{2\rho\sigma}\nabla_\nu h^2_{\sigma\mu}\big)+\frac12\nabla_\rho\big( h^{2\rho\sigma}\nabla_\sigma h^2_{\mu\nu}\big)-\frac12\nabla^\rho \nabla^\sigma h^2_{\rho\sigma} h^2_{\mu\nu}-\frac{216}{7L^4}s^2\nabla_\mu\nabla_\nu s^2\nn\\
&+\frac54 h^{2\rho\sigma}\nabla_\nu \nabla_\mu h^2_{\rho\sigma}+\frac14\nabla^\rho(h^2+\phi^2)\big(\nabla_\mu h^2_{\rho\nu}+\nabla_\nu h^2_{\rho\mu}\big)-\frac14\nabla^\rho(h^2+\phi^2)\nabla_\rho h^2_{\mu\nu}+\frac{23}{L^2}h^2h^2_{\mu\nu}\nn\\
&-\frac12\nabla^\sigma h^2_{\rho\nu}\nabla^\rho h^2_{\sigma\mu}+\frac12\nabla^\sigma h^{2\rho}_{\nu}\nabla_\sigma h^2_{\mu\rho}+\frac{8}{L^2} h^{2\rho}_{\nu}h^2_{\mu\rho}+\frac{17}{2L^2}g_{\mu\nu}h^{2\rho\sigma}h^2_{\rho\sigma}+\frac12g_{\mu\nu}\nabla_\rho\big( h^{2\rho\sigma}\nabla_\sigma \phi^2\big)\nn\\
&+{\frac3{16}}g_{\mu\nu}\nabla^\rho h^2\nabla_\rho h^2{-\frac12}g_{\mu\nu}\nabla^\rho h^2\nabla_\rho \phi^2+{\frac{5}{112}}g_{\mu\nu}\nabla^\rho \phi^2\nabla_\rho \phi^2-\frac{201}{98L^2}g_{\mu\nu}\phi^2\phi^2-\frac{40}{7L^2}g_{\mu\nu}\phi^2h^2\nn\\
&-\frac{33}{L^2}g_{\mu\nu}h^2h^2-\frac{1}{7L^2}h^2_{\mu\nu}\phi^2+\frac{432}{7L^4}h^2_{\mu\nu} s^2+\frac12h^2_{\mu\nu}\square h^2+\frac34g_{\mu\nu}h^{2\rho\sigma}\nabla^\tau\nabla_\rho h^2_{\tau\sigma}+\frac{26352}{49L^6}g_{\mu\nu}s^2 s^2\nn\\
&-\frac{108}{49L^4}g_{\mu\nu}\phi^2s^2-\frac{11}{6L^2}h^2_{\mu\nu}u^2-{\frac{1}{18L^2}}g_{\mu\nu}u^2u^2+{\frac{25}{8L^2}g_{\mu\nu}h^2u^2}-\frac1{2}g_{\mu\nu}\Big(\frac34h^{2\rho\sigma}\nabla_\rho \nabla_\sigma h^2\nn\\
&-\frac12\nabla^\rho \nabla^\sigma h^2_{\rho\sigma} h^2+\frac54 h^{2\rho\sigma} \nabla_\rho\nabla_\sigma \phi^2-\frac34 \nabla^\lambda h^{2\rho\sigma}\nabla_\lambda h^2_{\rho\sigma}+\frac{216}{7L^4}\nabla^\rho s^2\nabla_\rho s^2{+\frac12}h^2\square h^2\nn\\
&+\frac12\nabla^\rho\phi^2\nabla^\sigma h^2_{\rho\sigma}-{\frac{46}{56L^2}}\phi^2u^2+\frac{216}{7L^4}h^2s^2-\frac{36}{7L^4}u^2s^2\Big)-\frac1{48}g_{\mu\nu}\Big(\frac1{2}\square h^2u^2+\nabla^\lambda  h^2\nabla_\lambda  u^2\nn\\
&-\nabla^\lambda \phi^2\nabla_\lambda  u^2
\Big)\Big\}+\frac{1}{48}\big(\nabla_{\mu}\nabla_{\nu}t_+^1t_+^1+\frac12\nabla_{\mu} t_+^1\nabla_{\nu} t_+^1\big)+\frac{1}{96}g_{\mu\nu}\big(\nabla_\rho t_+^1\nabla^\rho t_+^1-\frac{16}{L^2}t_+^1t_+^1\big)=0.
\end{align}  
\endgroup
In our case, the fluctuation modes $(h^{I_{1}}_{\mu\nu}$, $\phi^{I_{1}}$, etc) represent the deviations of the LLM geometry from the AdS$_{4}\times S^7$ space. In that case, we have shown that the symmetrized-traceless transverse KK graviton mode $\check h^2_{(\mu\nu)}$, which is given by~\cite{Jang:2016tbk} 
\begin{align}
\check h_{(\mu\nu)}^{2}
=
\hat\phi^{2}_{(\mu\nu)}
+\frac7{30}\hat\psi^{2}_{(\mu\nu)}
+\frac{L^2}{8}\nabla_{(\mu}\nabla_{\nu)}\hat\phi^{2}
-\frac{7L^2}{720}\nabla_{(\mu}\nabla_{\nu)}\hat\psi^{2},
\end{align}
with $\hat\psi_{\mu\nu}^{2}\equiv 18h^{2}_{\mu\nu}-\frac{L}{2}(\nabla_{\mu}t^{2}_{\nu}+\nabla_{\nu}t^{2}_{\mu})$, $\hat\phi_{\mu\nu}^{2}\equiv -\frac75\left(h_{\mu\nu}^{2}+\nabla_{\mu}\nabla_{\nu}s^{2}\right)$, $\hat\phi^2\equiv \phi^2+\frac{16}{L^2}s^2$, is vanishing at linear order in $\mu_0$. This implies, at linear order in $\mu_0$, $h^2_{\mu\nu}$ is not an independent tensor mode and can be expressed in terms of the scalar modes as
\begin{align}\label{hmn-ans}
 h_{\mu\nu}^{2}
=&-\frac{L^2}{576}\nabla_\mu\nabla_\nu \hat\psi^2+\frac{11L^2}{224}\nabla_\mu\nabla_\nu \phi^2-\frac{3}{14}\nabla_\mu\nabla_\nu s^2-g_{\mu\nu}\Big(\frac1{72}\hat\psi^2+\frac3{28}\hat\phi^2\Big). 
\end{align}
We have also shown that, only two gauge invariant combinations of the four scalar fields $h^2,\phi^2, u^2, s^2$ are physical modes.  The two gauge invariant physical scalar modes are $\hat \psi^2$ and $\hat\phi^2$  so that we express  the four scalars  as
\begin{align}\label{GF}
h^2=a_1\hat\psi^2+b_1\hat\phi^2,\quad u^2=a_2\hat\psi^2+b_2\hat\phi^2,\quad \phi^2=a_3\hat\psi^2+b_3\hat\phi^2 ,\quad s^2=a_4\hat\psi^2+b_4\hat\phi^2.
\end{align}
In particular, these relations should be valid when we substitute the values of those scalar fields from the asymptotic expansion of the LLM solutions. In that case we find the following relations among  the constants $a_i$ and $b_i$,
\begin{align}\label{GF0}
b_1=-\frac37(1-6a_1),\quad b_2=-\frac{18}{7}(2-a_2),\quad b_3=\frac37(1+6a_3),\quad b_4=\frac{L^2}{28}\Big(1+\frac{72a_4}{L^2}\Big).
\end{align}
It is more convenient to use the diagonal modes $\check\phi^2=\frac9{70}(7\hat\psi^2+18\hat\phi^2),~\check\psi^2=\frac1{70}(7\hat\psi^2-162\hat\phi^2)$.  See \cite{Jang:2016tbk} for the derivation of these diagonal modes. Then we can write 
\begin{align}\label{GF1}
&h^2=\frac16\check\psi^2+\frac19\Big(10a_1-\frac16\Big)\check\phi^2,\quad u^2=2\check\psi^2+\frac29\Big(5a_2-1\Big)\check\phi^2,\nn\\
&\phi^2=-\frac16\check\psi^2+\frac19\Big(10a_3+\frac16\Big)\check\phi^2 ,\quad s^2=-\frac{L^2}{72}\check\psi^2+\frac19\Big(10a_4+\frac{L^2}{72}\Big)\check\phi^2.
\end{align}
In the LLM solution, the asymptotic expansion  of  $\check\phi^{2}$ is of order $\mu_0^3$  and it does not contribute to the equations of motion at $\mu_0^2$ order.  Therefore, we can set $\check\phi^{2}$ to zero and use \eqref{hmn-ans} and \eqref{GF1} in \eqref{lineq1-3-06} to write the quadratic part of the equation of motion only in terms of $\check\psi^2$ as follows
\begin{align}\label{lineq1-3-06a}
&\Big(L_E+\frac{12}{L^2} \Big)\hat h_{\mu\nu}^0+\frac1{34560}\Big\{-\frac{26}3\nabla_\mu\check\psi^2\nabla_\nu\check\psi^2+\frac{28}3\check\psi^2\nabla_\mu\nabla_\nu\check\psi^2+\frac{L^2}3\nabla_\mu\nabla^\rho\check\psi^2\nabla_\nu\nabla_\rho\check\psi^2\nn\\
&+\frac{L^2}2\nabla^\rho\check\psi^2\nabla_\mu\nabla_\nu\nabla_\rho\check\psi^2+\frac{L^4}{24}\nabla_\mu\nabla^\rho\nabla^\sigma\check\psi^2\nabla_\nu\nabla_\rho\nabla_\sigma\check\psi^2+\frac{L^4}{32}\nabla^\rho\nabla^\sigma\check\psi^2\nabla_\mu\nabla_\nu\nabla_\rho\nabla_\sigma\check\psi^2\nn\\
&-g_{\mu\nu}\Big(\frac{12}{L^2}\check\psi^2\check\psi^2+\nabla^\rho\check\psi^2\nabla_\rho\check\psi^2+\frac{35L^2}{48}\nabla^\rho\nabla^\sigma\check\psi^2\nabla_\rho\nabla_\sigma\check\psi^2-\frac{L^4}{64}\nabla^\tau\nabla^\rho\nabla^\sigma\check\psi^2\nabla_\tau\nabla_\rho\nabla_\sigma\check\psi^2\Big)\Big\}\nn\\
&+\frac{1}{48}\big(\nabla_{\mu}\nabla_{\nu} t_+^1t_+^1+\frac12\nabla_{\mu} t_+^1\nabla_{\nu} t_+^1\big)
+\frac{1}{96}g_{\mu\nu}\big(\nabla_\rho t_+^{1}\nabla^\rho t_+^1-\frac{16}{L^2} t_+^1 t_+^1\big)=0.
\end{align}

In the next subsection we introduce some non-trivial field redefinitions in order to eliminate the higher derivative terms in the above equation and obtain a linearized equation for the 4-dimensional fields.

\subsection {The KK mapping at quadratic order}

Our goal in this section is to apply the KK reduction procedure to the 11-dimensional supergravity on LLM geometry
 and construct 4-dimensional gravity theory whose solution encodes the information about the asymptotic limit of the LLM geometry.
In the previous subsection, using the graviton mode example in \eqref{lineq1-3-06a}, we have shown that the compactification of 11-dimensional supergravity on $S^7$ results in the field equations which contain higher derivative terms.
In general, the same is true for all the modes in the KK towers. To absorb the higher derivative terms we need to introduce some field redefinitions.
For instance, for some scalar KK mode whose equation of motion contains up to four derivatives, the field redefinition is of the form
\begin{align}\label{field-red1}
 S^{I}  =  s^{I}+K_{IJ_1J_2} t^{J_1} t^{J_2}
+L_{IJ_1J_2}\nabla_\mu  t^{J_1}\nabla^\mu t^{J_2},
\end{align}
where $K_{IJ_1J_2}$, $L_{IJ_1J_2}$ are some numerical coefficients, $s^{I}$ represents a gauge invariant 11-dimensional field and $S^{I}$ is the corresponding 4-dimensional field. See \cite{Skenderis:2006uy} for a systematic procedure of the KK reduction. The $t^{J_i}$'s represent all the fields that appear in higher derivative part of the equations of motion of $s^I$.
If the field equation contains more than four derivatives, then the field redefinition will contain more than two derivatives. For the important scalar modes $\check\psi^2$ and $t^1_+$ discussed in the previous subsection, the equations of motion involve higher derivative terms only if we want to solve them at cubic or higher order in  $\mu_0$.  Up to quadratic order, their equations of motion are linear and are given by 
\begin{align}\label{diagpp}
\left(\square-M^2_t\right)t_+^{1}=0,\qquad\left(\square-M^2_\psi \right)\check\psi{^2}=0,
\end{align}
where 
$M^2_t=M^2_\psi=-\frac{8}{L^2}$. In this case, the field redefinitions are trivial and we can write the corresponding 4-dimensional fields as, 
\begin{align}\label{PsiT}
\Psi=\check\psi^2
,\qquad
T=t^1_+
.
\end{align}

The 4-dimensional gravity action with matter, which yields the equations of motion in \eqref{diagpp} for $\Psi$ and $T$ is given by 
\begin{align}\label{4dact}
S = \frac{1}{16\pi G_N^{(4)}}\int d^4 x \sqrt{-g}\left(\hat R - 2\Lambda\right) + S_m, 
\end{align}
where $\Lambda=-\frac{(d-1)(d-2)}{2L^2_{\rm AdS_4}}=-\frac{12}{L^2}$ and 
\begin{align}\label{S_m}
S_m=&-\frac{A_{t}}{2}\int d^4x\sqrt{-g}(\nabla_\mu T\nabla^\mu T+M_{t}^2T^{2})-\frac{A_{\psi}}{2}\int d^4x\sqrt{-g}(\nabla_\mu \Psi\nabla^\mu \Psi+M_{\psi}^2\Psi^{2}),
\end{align} 
for some over all normalizations $A_{t}$ and $A_{\psi}$ which will be fixed later. The corresponding energy-momentum tensor is
\begin{align}\label{Tmn}
\tilde T_{\mu\nu}=-\frac2{\sqrt{-g}}\frac{\delta S_m}{\delta g^{\mu\nu}}&=A_t\Big[\nabla_\mu  T\nabla_\nu  T-\frac12g_{\mu\nu}\big(\nabla_\rho T\nabla^\rho  T+M_t^2 T^{2}\big)\Big]\nn\\
&+A_{\psi}\Big[\nabla_\mu \Psi\nabla_\nu \Psi-\frac12g_{\mu\nu}\big(\nabla_\rho \Psi\nabla^\rho \Psi+M_{\psi}^2\Psi^{2}\big)\Big].\end{align}

The next step is to obtain the equation of motion for the 4-dimensional graviton by using the result in \eqref{lineq1-3-06a} and the above energy-momentum tensor. Since the background is AdS$_4$, we consider  the 4-dimensional Einstein equation with negative cosmological constant 
\begin{align}
\hat R_{\mu\nu}-\frac12\hat g_{\mu\nu}\hat R+\Lambda \hat g_{\mu\nu}=8\pi G_N  \tilde T_{\mu\nu},\label{4DG}
\end{align}
where $\hat g_{\mu\nu}=g_{\mu\nu}+\delta g_{\mu\nu}$ represents the metric which is deviated from the AdS$_4$ due to the presence of the energy-momentum tensor of the matter fields $\Psi$ and $T$. In order to obtain the equations for the graviton field, we insert the perturbed metric into \eqref{4DG} and keep only up to the terms that are linear in the fluctuations 
\begin{align}
\delta R_{\mu\nu}-\frac12\delta g_{\mu\nu}R-\frac12g_{\mu\nu}\delta g^{\rho\sigma}R_{\rho\sigma}-\frac12g_{\mu\nu}g^{\rho\sigma}\delta R_{\rho\sigma}-\frac{12}{L^2} \delta g_{\mu\nu}=8\pi G_N \tilde T_{\mu\nu}.
\end{align}
Setting $\delta g_{\mu\nu}= H_{\mu\nu},~\delta g^{\mu\nu}=-H^{\mu\nu}$, we can write
\begin{align}\label{Hmn-eq}
&\frac12\Big(-\square  H_{\mu\nu}+\nabla^\rho\nabla_\mu  H_{\rho\nu}+\nabla^\rho\nabla_\nu  H_{\rho\mu}-\nabla_{\mu}\nabla_{\nu} H\Big)+\frac{12}{L^2} H_{\mu\nu}-\frac{6}{L^2}g_{\mu\nu} H
\nn\\
-&\frac12g_{\mu\nu}(\nabla^\rho\nabla^\sigma H_{\rho\sigma}-\square H)=8\pi G_N \tilde T_{\mu\nu}.\end{align}
Taking the trace, we obtain
\begin{align}
-\frac{12}{L^2} H-\nabla^\rho\nabla^\sigma H_{\rho\sigma}+\square H=8\pi G_Ng^{\rho\sigma} \tilde T_{\rho\sigma}.\end{align}
Plugging this into \eqref{Hmn-eq} and using the energy-momentum tensor in \eqref{Tmn}, we obtain
\begin{align}\label{Hmn-eq4}
\Big(L_E  +\frac{12}{L^2}\Big)H_{\mu\nu}&-8\pi G_N A_{t}\Big( \nabla_\mu T\nabla_\nu T+\frac{M_t^2}2g_{\mu\nu}T^{2}\Big)\nn\\
&-8\pi G_N A_{\psi}\Big( \nabla_\mu \Psi\nabla_\nu \Psi+\frac{M_{\psi}^2}2g_{\mu\nu}\Psi^{2}\Big)=0.\end{align}

The final step is to introduce a field redefinition which eliminates the higher derivative terms in \eqref{lineq1-3-06a} and reduces it to \eqref{Hmn-eq4}.  Since the equation of motion in \eqref{lineq1-3-06a} contains up to six derivative terms in $\check \psi^{2}$ and no higher derivative term in $t^1_+$, the field redefinition should contain up to four derivative terms in $\check \psi^{2}$ and no derivative in $t^1_+$. The required field redefinition is given by
\begin{align}\label{FRD1}
H_{\mu\nu}&=\hat h^{0}_{\mu\nu}+g_{\mu\nu}\big(\tilde C_1\check\psi^2\check\psi^2+\tilde C_2\nabla^\rho\check\psi^2\nabla_\rho\check\psi^2\big)+\tilde C_3\nabla_\mu\check\psi^2\nabla_\nu\check\psi^2\nn\\
&~~+g_{\mu\nu}\tilde C_4\nabla^\rho\nabla^\sigma\check\psi^2\nabla_\rho\nabla_\sigma\check\psi^2+\tilde C_5\nabla_\mu\nabla^\rho\check\psi^2\nabla_\nu\nabla_\rho\check\psi^2+g_{\mu\nu}\tilde C_t  t^{1}_+ t^{1}_+.
\end{align}
Inserting  \eqref{FRD1} into \eqref{Hmn-eq4} and comparing the result with \eqref{lineq1-3-06a}, we determine the unknown  coefficients listed below,
\begin{align}\label{coefs}
& \tilde C_1=-\frac{1}{2^6\, 3^3\, 5},\quad \tilde C_2=-\frac{L^2}{2^{11}\,3^3\,5},\quad \tilde C_3=-\frac{7L^2}{2^{11}\,3^4\,5},\quad  \tilde C_4=-\frac{L^4}{2^{14}\,3^3\,5},\\
& \tilde C_5=-\frac{L^4}{2^{13}\,3^4\,5},\quad \tilde C_t=-\frac{1}{2^5\,3},\quad  8\pi G_NA_t=\frac{1}{2^5\,3},\quad \quad 8\pi G_NA_{\psi}=\frac{1}{2^8\,3^2}.\nn
\end{align}
In this paper we focused on obtaining the field redefinition for the graviton mode. A similar procedure determines the field redefinition for higher KK tensor, vector, and scalar modes in 4-dimension. In \cite{Skenderis:2006uy,Skenderis:2006di} a similar result was obtained for 5-dimensional KK towers, which are obtained from the dimensional reduction of the 10-dimensional type-IIB supergravity on background which is asymptotically AdS${}_{5}\times S^{5}$. The resulting field redefinition was dubbed the KK map between the 10-dimensional and 5-dimensional fields. 
In our case, the KK map for graviton field is as in \eqref{FRD1}.

\section{HEE from \textbf{\textit{Vev}} and \textbf{\textit{Source}}}

When the conformal symmetry is broken due to a relevant deformation of a CFT, the CPOs develop non-vanishing one-point functions.
In that case, one can infer that the deviation of the EE from it's value in the CFT is related to the non-vanishing one-point function.
From the perspective of the dual gravity, the non-conformal field theory corresponds to a gravity theory on an asymptotically AdS geometry.
According to the RT conjecture, the deviation of the geometry from the AdS space induces a variation in HEE.
In other words, non-vanishing one-point function of CPO of gauge theory induces the variation of EE and induces change of geometry in dual gravity.
The dynamics of the geometry is governed by the linearized Einstein equations with matter fields interactions on the AdS background. Understanding the field theory counterpart of  these governing equations is intriguing. Research in this direction is progressing. In this paper, we move further this progress by studying the HEE in mABJM theory.

In \cite{Jang:2016tbk}, we presented the calculation of one-point function of the CPO with conformal dimension one as an evidence supporting the duality between the mABJM theory and 11-dimensional supergravity theory on the LLM geometries. 
We obtained the one-point function, $\langle {\cal O}^{(1)}\rangle$  in the large $N$ limit, using the KK holography method from the 11-dimensional LLM geometry and showed that there is an exact agreement with the results obtained by using the supersymmetric vacua of mABJM theory. We conducted this test for all supersymmetric vacua, which are infinite in number.
In order to strengthen the confirmation of the duality of the two theories, those results will be extended to the case of conformal dimension two CPO \cite{Jang:2018aqr}. In this paper, we exploit this exact correspondence to study the HEE based on the RT conjecture. 
To that end, we start by reading the asymptotically AdS solutions of the 4-dimensional gravity theory, which are constructed in the previous section, from the 11-dimensional LLM solutions. Applying the RT conjecture to those 4-dimensional gravity solutions, we calculate the leading order deviation of the HEE ($\delta S$) from its value in pure AdS$_4$ space.  We compare the result with the known result of the HEE in the LLM geometry \cite{Kim:2016dzw}.

\subsection{Vacua of mABJM theory and LLM geometries}
Under the supersymmetry preserving mass deformation of the ABJM theory, the global SU(4) symmetry of the ABJM theory is broken to SU(2)$\times$SU(2)$\times$U(1). To express the vacuum solution which reflects the broken symmetry, we split the scalar fields into $Y^A = (Z^a,\, W^{\dagger a})$, where $A = 1,2,3,4$ and $a = 1,2$. One interesting feature of the mABJM theory is that it has discrete Higgs vacua, which are represented as direct sums of GRVV matrices~\cite{Gomis:2008vc}. The vacua are classified by occupation numbers, $N_n$ and $N_n'$ which are respectively denote the numbers of $n\times (n+1)$ GRVV matrices ${\cal M}_n$ and $(n+1)\times n$ GRVV matrices $\bar {\cal M}_n$, in the direct sums. 
See \cite{Cheon:2011gv} for the details.  
These vacua are supersymmetric if the occupation numbers are in the range, $0\le N_n,\, N_n'\le k$~\cite{Kim:2010mr}. There is a one-to-one map between the discrete Higgs vacua of the mABJM theory and the LLM geometries with ${\mathbb Z}_k$ orbifold, which have SO(2,1)$\times {\rm SO}(4)/{\mathbb Z}_k \times {\rm SO}(4)/{\mathbb Z}_k$ isometry.

 The LLM  metric and the corresponding 4-form field strength are given by
\begin{align}\label{LLMds2}
ds^2 &= -{\bf G}_{tt} ( -dt^2 + dw_1^2 + dw_2^2) + {\bf G}_{xx} (d\tilde x^2 + d\tilde y^2) 
+ {\bf G}_{\theta\theta} ds^2_{S^3/\mathbb{Z}_k}
+ {\bf G}_{\tilde\theta\tilde\theta} ds^2_{\tilde S^3/\mathbb{Z}_k},\nn\\
{\bf F}_4 &= -d \left(e^{2\Phi}h^{-2}V \right)
\wedge dt\wedge dw_1\wedge dw_2 +\mu_0^{-1} \left[Vd(\tilde y^2e^{2G}) + h^2e^{3G}\star_2 d(\tilde y^2 e^{-2G})\right]
\wedge d\Omega_3
\nn \\
&~~~+\mu_0^{-1}
\left[ Vd(\tilde y^2e^{-2G}) -h^2e^{-3G}\star_2 d(\tilde y^2 e^{2G})\right]
\wedge d\tilde\Omega_3,
\end{align}
where $\mu_0$ is a mass parameter, $ds^2_{S^3/\mathbb{Z}_k}$ and $ds^2_{\tilde S^3/\mathbb{Z}_k}$ are line elements of three-spheres with ${\mathbb Z}_k$ orbifold, and 
 $d\Omega_3=-(\sin\theta/8)d\theta\wedge d\phi\wedge d\psi$, $d\tilde\Omega_3=-(\sin\tilde\theta/8)d\tilde\theta\wedge d\tilde\phi\wedge d\tilde\psi$  are the volume forms of the two spheres in the Euler coordinate system.
The warp factors and the functions defining the 4-form field strength are given by
\begin{align}\label{warpfac}
&{\bf G}_{tt}= -\left(\frac{4\mu_0^2 \tilde y\sqrt{\frac14 - Z^2}}{f^2}
\right)^{2/3},\quad 
{\bf G}_{xx}=\left(\frac{f\sqrt{\frac14 - Z^2}}{2 \mu_0\tilde{y}^2}\right)^{2/3},
\quad {\bf G}_{\theta\theta} =\left(
\frac{f\tilde y \sqrt{\frac12 + Z}}{2 \mu_0\left(\frac12 -Z\right)} \right)^{2/3},\nn\\
&{\bf G}_{\tilde\theta\tilde\theta}=\left( \frac{f\tilde y \sqrt{\frac12 - Z}}{2\mu_0 
\left(\frac12 + Z\right)}\right)^{2/3},\quad h^2=\frac{\sqrt{\frac14 - Z^2}}{\tilde y},
\quad e^{2\Phi}=\frac{4 \tilde y \mu_0^2 \sqrt{\frac14 - Z^2}}{f^2},\quad e^{2 G}=
\frac{\frac12 + Z}{\frac12 - Z}
\end{align}
with $f(\tilde x,\tilde y) = \sqrt{1 - 4 Z^2 - 4\tilde y^2 V^2}.$

Here we notice that the LLM geometry is completely determined by 
two functions, 
\begin{align}\label{ZandV}
Z(\tilde x,\tilde y)
=\sum_{i=1}^{2N_B\!+\!1}\frac{(-1)^{i\!+\!1}
(\tilde x\!-\!\tilde x_i)}{2\sqrt{(\tilde x\!-\!\tilde x_i)^2+\tilde y^2}}
\ ,\qquad
V(\tilde x,\tilde y)
=\sum_{i=1}^{2N_B\!+\!1}\frac{(-1)^{i\!+\!1}}{2\sqrt{(\tilde x\!-\!\tilde x_i)^2+\tilde y^2}},
\end{align}
where $\tilde x_i$'s are numerical parameters. 
The function $Z(\tilde x,\tilde y)$ at $\tilde y=0$ has a value $\frac12$ for  $\tilde x_{2i-1}< \tilde x < \tilde x_{2i}$ and it is $-\frac12$ for  $\tilde x_{2i}< \tilde x < \tilde x_{2i+1}$. The geometries are classified by those values of $Z(\tilde x,0)$.
That is, the LLM geometries are represented as an infinite strip in the $\tilde x$-direction with $Z(\tilde x,0)=-\frac12$ denoted by black strip and $Z(\tilde x,0)=\frac12$ denoted by white strip.
This representation of the LLM geometry is called the droplet picture. 
So the $\tilde x_i$'s denote the positions of the boundaries between the black and the white strips and $N_B$ is the number of finite-sized black/white strips in the droplet representation. 
Due to flux quantization condition of the 4-form field strength, the difference between consecutive $\tilde x_i$'s is quantized as \cite{Cheon:2011gv}
\begin{align}
\tilde x_{i+1}-\tilde x_i = 2\pi l_{{\rm P}}^3\mu_0 {\mathbb Z},
\end{align}
where $l_{{\rm P}}$ is the Planck length. 
Therefore, all LLM geometries are completely determined by these quantized locus $\tilde x_i$'s.

In order to consider the gauge/gravity duality near the UV fixed point of the field theory, we need to expand the dual geometry in the asymptotic region. For the asymptotic expansion of the general LLM geometries, it is convenient to introduce new parameters~\cite{Kim:2016dzw}, 
\begin{align}
C_p = \sum_{i=1}^{\infty}(-1)^{i+1} \left( \frac{\tilde x_i}{2\pi l_{{\rm P}}^3 \mu_{0}\sqrt{A}}\right)^p,
\end{align} 
where $A$ is defined as\footnote{In \cite{Lin:2004nb}, there is also an alternative representation of the LLM solutions in terms of the Young diagrams. In that case, $A$ denotes the area of the Young diagram.}

\begin{align}\label{AA}
A = k N -\frac12 \sum_{n=0}^{\infty} \left[l_n (k-l_n) + l_n' (k-l_n')\right]. 
\end{align}
Here we introduce a new set of parameters, $\{ l_n, \,l_n'\}$, which are called discrete torsions and used to classify the LLM geometries in the droplet picture. See \cite{Cheon:2011gv} for the details. The one-to-one correspondence between the vacua of the mABJM theory and the LLM geometries identifies $\{N_n, N_n'\}$ with $\{l_n, l_n'\}$.

\subsection{KK reduction of the LLM geometries}\label{ast-AdS-metric}
In the Fefferman-Graham coordinate system, the LLM metric is given by (see  \cite{Jang:2016tbk} for details)
\begin{align}\label{dsFG}
ds^2 =&\frac{ L^2}{4 z^2}\left[dz^2 +\frac{4z^2}{L^2} [1+{\tilde g}_1(z,\tau )]\left( -dt^2+dw_1^2+dw_2^2\right) \right]\nn\\
&+ [1+{\tilde g}_2(z,\tau ) ]d\tau^2+ [1+{\tilde g}_3(z,\tau )] ds_{S^3}^2 + [1+{\tilde g}_4(z,\tau  )] ds_{\tilde S^3}^2,
\end{align}
where the $\tilde g_i(z,\tau)$ represents the deviation of the LLM metric from the AdS$_4\times S^7$ background. Similarly, the 4-form field strength can be split into the background and the rest.  In the asymptotic region, these deviations become small fluctuations, and the LLM solution can be written as in \eqref{fluct}, where the values of the small fluctuations $h_{pq}$ and $f_{pqrs}$ are read from the asymptotic expansion of $\tilde g_i(z,\tau)$, and the similar terms in the 4-form field strength.  Then, these small fluctuations can be expanded in terms of the spherical harmonics on $S^7$, in order to obtain the values of the KK modes ($h^{I_1}_{\mu\nu}, \phi^{I_1}, ~ etc$) of the previous section.
In \cite{Jang:2016tbk}, we have listed the full result for all the KK modes up to $\mu_0^2$ order. Here, we are interested only in the graviton mode,  which corresponds to a combination of the KK zero modes. We also need ($I_1=2, I_{35}=1$) modes, which appear in the quadratic part of the graviton equation of motion. Therefore, we copy the following results from \cite{Jang:2016tbk}
\begin{align}\label{hhppp}
&h_{ij}^0 = \left[\frac{L^2\mu_0^2}{720}\left(360 + 7 \beta_3^2 \right) + {\cal O}\left(\mu_0^4\right)\right]\eta_{ij},\qquad h_{zz}^0 = 0, 
\nn \\ 
&h^0 = \frac{(\mu_0 z)^2}{60}\left(360 + 7 \beta_3^2\right) + {\cal O}\left(\mu_0^4\right), 
\qquad
\phi^0 = \frac{(\mu_0 z)^2}{15}\left(-80 + \beta_3^2\right) + {\cal O}\left(\mu_0^4\right), 
\nn \\
&\hat\psi^0 = \frac{6(\mu_0 z)^2}{5}\left( 80 + \beta_3^2\right)+ {\cal O}(\mu_0^4),
\qquad \check\psi^{2} = -24\beta_3\mu_0 z+{\cal O}(\mu_0^3),
\nn \\
&t_+^1 = 16\sqrt{3} \, \mu_0 z + {\cal O}(\mu_0^3),
\end{align}
where $\eta_{ij} = {\rm diag}(-1,1,1)$ and
\begin{align}\label{beta3}
\beta_3 = 2 C_1^3 - 3  C_1 C_2 +  C_3. 
\end{align}

In the previous section, we have established the KK maps which relate the above 11-dimensional KK modes to the corresponding canonical 4-dimensional gravity fields. These maps are given in \eqref{PsiT} and \eqref{FRD1}. Using these maps, we can write the asymptotically AdS$_4$ solution to the 4-dimensional gravity equations \eqref{diagpp} and \eqref{Hmn-eq4} from the KK reduction  of the LLM solution. The results are 

\begin{align}\label{Hmn}
&H_{ij} = \left[-\frac{(L\mu_0)^2}{180}\left( 30 + \beta_3^2\right)+{\cal O} \left(\mu_0^4\right)\right]\eta_{ij} , 
\qquad
H_{zz} = - \frac{(L\mu_0)^2}{1440}\left(960 + 29\beta_3^2\right) + {\cal O}\left(\mu_0^4\right)\nn\\
&\Psi = -24 \beta_3 \mu_0 z + {\cal O}(\mu_0^3), \qquad T = 16\sqrt{3} \, \mu_0 z + {\cal O}(\mu_0^3).\end{align}

\subsection{One-point function for the CPO with $\Delta=1$}

The $vev$ of a CPO with conformal dimension one in mABJM was obtained in \cite{Jang:2016tbk}. For clarity of presentation, we shortly review that result here. The CPO in ABJM theory with conformal dimension one, which preserves the SU(2)$\times$SU(2)$\times$U(1) global symmetry and has non-vanishing $vev$, is given by
\begin{align}\label{CPO}
{\cal O}^{(1)} = \frac1{2\sqrt{2}} {\rm Tr}\left( Z^a Z_a^\dagger - W^{\dagger a} W_a\right).
\end{align}
Since the supersymmetry of the ABJM theory protects the scalar fields from quantum corrections and the contributions from the multi-trace terms are suppressed by $1/N$ as compared to single-trace terms, the $vevs$ of the CPOs are exactly determined by the classical values for scalar fields in the large $N$ limit~\cite{Jang:2016tbk}. 
Based on this argument, we obtained the $vev$ of the CPO with conformal dimension one for all supersymmetric vacua in large $N$ limit,
\begin{align}\label{vevCPO2}
\langle {\cal O}^{(1)}\rangle_m = \frac{k \mu }{4\sqrt{2}\, \pi } 
\sum_{n=0}^{\infty}n(n+1)(N_n - N_n'),
\end{align}
where $\langle\cdots\rangle_m$ represents the $vev$ in the mABJM theory and $\mu$ is the mass parameter related to the LLM geometry mass parameter as $\mu = 4\mu_0$.

In the gauge/gravity duality, the relation between the conformal dimension of gauge invariant operators and the mass of the dual scalar modes in the 4-dimensional gravity theory is given by 
\begin{align}\label{ScalarMass}
\frac{m_\phi^2 L^2}{4} = \Delta (\Delta -3).
\end{align}
In our case the gravity mode dual to the CPO with conformal dimension one is the mode $\Psi$ in \eqref{diagpp}. 
The gauge/gravity duality dictionary states that, the $vev$ of a CPO with conformal dimension $\Delta$ is determined by the coefficient of $z^\Delta$ in the asymptotic expansion of the dual scalar field. According to this rule,  the $vev$ of the CPO of conformal dimension one is determined by the asymptotic expansion of $\Psi$ in \eqref{Hmn} as
\begin{align}\label{dual_rel}
\langle {\cal O}^{(1)}\rangle_m
= -  \frac{24 N^2}{\sqrt{\lambda}}\, {\mathbb N} \, \beta_3 \mu_0,
\end{align}
where ${\mathbb N}$ depends on the normaliation of the scalar field $\Psi$ and $\lambda= N/k$ is the 't Hooft coupling constant in the ABJM theory. The overall factor $N^2/\sqrt{\lambda}$ is originated from the gauge/gravity dual relation in 4-dimensional gravity, 
\begin{align}
\frac1{16\pi G_N^{(4)}}\sim \frac{N^2}{\sqrt{\lambda} L^2}.
\end{align}
In order to fix the normalization ${\mathbb N}$,  we use the identity (see \cite{Jang:2016tbk} for the proof)
\begin{align}\label{beta3_2}
\beta_3= \frac{3}{A^{\frac32}}\sum_{n=0}^{N_B}n(n+1)(l_n - l_n'). 
\end{align}
For $k=1$ and the general $N_B$ or the general $k$ and $N_B=1$ cases, this identity is valid for any $N\ge 2$. However, for both $k$ and $N_B$ greater than one, the right-hand side of \eqref{beta3_2} is only the leading order term in the $\frac1N$-expansion of $\beta_3$. In the large $N$ limit, we note that the leading contribution of $A$ is $A = kN$. Therefore, recalling that the field theory result in \eqref{vevCPO2} is obtained in the large $N$ limit, we can fix the normalization as ${\mathbb N} = -\frac{\sqrt{2}}{144\pi}$ by comparing the field theory result and the gravity result \eqref{dual_rel} in the large $N$ limit.  The one-to-one map $\{l_n,\, l_n'\}
\Longleftrightarrow
\{N_n,\, N_n'\}$ is also used. Then the $vev$ can be written as 
\begin{align}\label{vevCPO3}
\langle {\cal O}^{(1)}\rangle_m =\frac{N^2\mu_0}{3\sqrt{2}\,\pi\sqrt{\lambda}}\, \beta_3.
\end{align}
In \cite{Jang:2016tbk}, we have verified that $\beta_3$ is independent of $N$, so that the overall normalization factor in \eqref{vevCPO3} is proportional to $N^{\frac32}$ in the case $k=1$, which is the well-known relation in M2-brane theory.  The above relation \eqref{vevCPO3} gives an exact dual relation in large $N$ limit for all supersymmetric vacua in mABJM theory and the corresponding LLM geometries with ${\mathbb Z}_k$ orbifold.

\subsection{HEE from  LLM geometries in 4-dimensions}
According to the RT conjecture, the HEE with a subspace $A$ at a fixed time on the boundary of ($d$+1)-dimensional AdS geometry is given by 
\begin{align}\label{S_A}
S_{A} = \frac{{\rm Min}(\gamma_A)}{4 G_N},
\end{align}
where $G_N$ is the Newton constant in the ($d$+1)-dimensional gravity theory and $\gamma_A$ is an area of the surface stretched to the bulk direction, which has the same boundary with the subsystem $A$.  The surface is expressed by the induced metric,
\begin{align}\label{indmet}
 g^{(0)}_{ij} = \frac{\partial w^\mu}{\partial \sigma^i}\frac{\partial w^\nu}{\partial \sigma^j} g_{\mu\nu}, 
\end{align}
where $\sigma^1,\cdots, \sigma^{d-1}$ are coordinates on the surface, $g_{\mu\nu}$ is the asymptotically AdS$_{d+1}$ bulk metric, and $w^0, w^1,\cdots, w^{d}$ are the bulk coordinates. 
The area $\gamma_A$ is given by 
\begin{align}\label{gamA}
\gamma_A = \int d^{d-1}\sigma \sqrt{\det \tilde g_{ij}}\,. 
\end{align}

In section \ref{4dgravity}, we have constructed a 4-dimensional gravity theory using KK reduction from the 11-dimensional gravity. We showed that, up to the quadratic order in the mass parameter $\mu_0$, there are only two scalar fields that are coupled to the 4-dimensional metric.  These two scalars carry the information of the asymptotic expansion of the LLM geometry. In the absence of these scalar fields, the geometry is pure AdS$_4$, and the scalar fields induce the deviation from the AdS$_4$ space. In this section, we use the asymptotically AdS$_4$ metric in subsection \ref{ast-AdS-metric} to compute the variation of the HEE $\delta S_A$ from its original value $S^0_A$ in pure AdS$_ 4$ geometry. 

\subsubsection{HEE from pure AdS geometry in 4-dimensions}
Before we proceed to the calculation of  $\delta S_A$, let us summarize the calculation of the HEE for the pure AdS$_4$ \cite{Ryu:2006ef}. The 4-dimensional AdS metric is given by 
\begin{align}
ds^2_{{\rm AdS}} = \frac{L_{{\rm AdS}}^2}{z^2}\left( -dt^2 + dw_1^2 + dw_2^2 + dz^2 \right),
\end{align}
where $L_{{\rm AdS}}$ denotes the radius of the AdS$_4$ geometry and $z$ represents the holographic direction. 
To obtain the HEE for a subspace $A$ which is a disk of radius $l$, we consider a mapping for the codimension 2 coordinates $\sigma^{1,2}$,
\begin{align}\label{map1}
t = {\rm constant}, \quad
w_1 = \sigma^2\cos\sigma^1, \quad
w_2 = \sigma^2\sin\sigma^1, \quad
z = z(\sigma^2). 
\end{align}
The components of the induced metric \eqref{indmet} are given by 
\begin{align}\label{tilgij2}
\tilde g_{11}^{(0)} = \frac{L_{{\rm AdS}}^2 \,\rho^2}{z^2}, 
\qquad
\tilde g_{12}^{(0)} = 0, \qquad
\tilde g_{22}^{(0)} = \frac{L_{{\rm AdS}}^2 }{z^2}\left(1 + z'^2\right), 
\end{align}
where we set $\sigma^2 = \rho$ and $z'\equiv  \left(\partial z/\partial\rho\right)$. Then the area of the surface $\gamma_A$ is given by 
\begin{align}\label{gamA2}
\gamma_A = \int_0^{2\pi}d\sigma^1\int_0^{l}d\rho \sqrt{\det\tilde g_{ij}} = 2\pi L_{{\rm AdS}}^2 \int_0^{l}d\rho \frac{\rho}{z^2}\sqrt{1+ z'^2}.
\end{align} 
The solution of $z(\rho)$ which minimizes $\gamma_A$ is
\begin{align}\label{solz}
z(\rho) = \sqrt{l^2 - \rho^2}
\end{align}
with boundary conditions 
\begin{align}
z(l) =0,\qquad z'(0) = 0. 
\end{align}
Computing the minimum area of $\gamma_A$, we obtain the HEE, 
\begin{align}
S_{A}^{(0)} = \frac{\pi L^2}{8 G_N^{(4)}}\left(\frac{l}{\epsilon} - 1\right),
\end{align}
where $\epsilon$ is the UV cut-off in the $z$-direction.

\subsection{Variation of HEE  from LLM geometries in 4-dimensions}
The asymptotically AdS$_4$ metric can be split into  the pure AdS part and fluctuations  as
 \begin{align}\label{4Dgmn}
\hat g_{\mu\nu} = g_{\mu\nu} + H_{\mu\nu},
\end{align}
where $g_{\mu\nu}$ is metric of the pure AdS$_4$. 
Then the induced metric \eqref{indmet} is given by 
\begin{align}\label{defmet}
\tilde g_{ij} &=  \frac{\partial w^\mu}{\partial \sigma^i}\frac{\partial w^\nu}{\partial \sigma^j} \hat g_{\mu\nu} = \frac{\partial w^\mu}{\partial \sigma^i}\frac{\partial w^\nu}{\partial \sigma^j}\left(g_{\mu\nu} + H_{\mu\nu}\right)
= \tilde g_{ij}^{(0)} + \tilde H_{ij},
\end{align}
where $\tilde g_{ij}^{(0)}$ is the induced AdS$_4$ metric and $ \tilde H_{ij}$
is determined from the asymptotic AdS$_4$ solution given in \eqref{Hmn}. Specifically,  
\begin{align}\label{indH}
\tilde H_{11} &= \rho^2 H_{11} = -\frac{( L\mu_0)^2\rho^2}{180}\left( 30 + \beta_3^2\right), 
\nn \\
\tilde H_{22} &= H_{11} + z'^2 H_{zz} = -\frac{(L\mu_0)^2}{1440}\left[ 240 + 8\beta_3^2 + \left(960 + 29 \beta_3^2\right) z'^2\right]. 
\end{align}

Apply the mapping \eqref{map1} to the induced metric \eqref{defmet}, we obtain the variation of the area, 
\begin{align}\label{delEE}
\delta \gamma_A &=\frac{1}{2}\int d^2\sigma \sqrt{\det\tilde g^{(0)}}\, \tilde g^{(0)ij}\tilde H_{ij}=\pi\int_0^l d\rho \sqrt{\det\tilde g^{(0)}}\, \tilde g^{(0)ij}\tilde H_{ij}.
\end{align}
Inserting \eqref{indH} into \eqref{delEE}, we obtain,
\begin{align}\label{gammaA1}
\delta\gamma_A &= -\frac{\pi L^2 \mu_0^2}{1440}\int_0^l d\rho \frac{\rho}{\sqrt{1 + z'^2}}\left[\left(1200 + 37\beta_3^2\right) z'^2 + 16 \left(30 + \beta_3^2\right)\right]
\nn \\
&= -\frac{\pi L^2 (\mu_0 l)^2}{48}\left(32 + \beta_3^2\right).
\end{align}
In the last step of \eqref{gammaA1}, we have used the solution of $z(\rho)$ given in \eqref{solz}. 
Therefore, the HEE up to $\mu_0^2$-order is given by 
\begin{align}\label{4dHEE}
S_A = S_A^{(0)} + \delta S_A &= \frac{\pi L^2}{8 G_N^{(4)}}\left[\frac{l}{\epsilon} - 1 - \frac43\left(1+ \frac{\beta_3^2}{32}\right)\left(\mu_0 l\right)^2\right]
\nn \\
&= \frac{\sqrt{2}\,\pi N^2}{3\sqrt{\lambda}}\left[\frac{l}{\epsilon} - 1 - \frac43\left(1+ \frac{\beta_3^2}{32}\right)\left(\mu_0 l\right)^2\right],
\end{align}
where the KK reduction relates the 4-dimensional and the 11-dimensional Newton's constant as
\begin{align}
\frac1{G_N^{(4)}} = \frac{{\rm vol}(S^7/{\mathbb Z}_k)}{G_N^{(11)}} = \frac{\pi^4 L^7}{3k\, G_N^{(11)}}= \frac{8\sqrt{2}\, N^2}{3\sqrt{\lambda}\, L^2}. 
\end{align}
Here we used the gauge/gravity dual relation in the ABJM theory and the 11-dimensional supergravity, $16\pi G_N^{(11)} = (2\pi)^8 l_{{\rm P}}^9$ and $L^6 = 32\pi^2 k N l_{{\rm P}}^6$. 
Inserting this into \eqref{4dHEE}, we reproduce the HEE from the LLM geometries in 11-dimensions \cite{Kim:2016dzw}.

As shown in section 2, the matter fields in the 4-dimensional gravity theory are determined by the 11-dimensional geometry. The equality of the HEE obtained from pure geometrical 11-dimensional gravity theory and the one obtained from the 4-dimensional matter-gravity theory shows that the information of the LLM geometry in the asymptotic limit is exactly encoded in the solutions of the matter fields in 4-dimensions.  Reversing this statement, the RT formula may play some role in the construction of geometrical solutions in higher dimensional theories from the solutions of matter fields in lower dimensional theories. It is intriguing to examine this possibility in our setup. However, it is highly non-trivial to achieve this goal because one has to extract the information of metric, which is local, from the HEE, which is non-local.

\subsection{HEE from holographic mapping of  \textbf{\textit {vev}} and \textbf{\textit{source}}}

In \cite{Jang:2016tbk}, we have discussed the gauge/gravity maps between gauge invariant operators in ABJM theory and scalar modes in 11-dimensional supergravity, which encode the information of LLM geometries with $S^7/{\mathbb Z}_k$ orbifold. Among the five infinite KK towers of scalar modes listed in \cite{Jang:2016tbk}, only the two scalar modes in \eqref{PsiT} are nonvanishing in the leading order of the asymptotic expansion of the LLM geometries. Using the relations between mass and conformal dimension listed in \cite{Jang:2016tbk}, we observe that the scalar field $\Psi$ with $M^2_{\Psi}=\frac{I(I-6)}{L^2}\big|_{I=2}$ is dual to CPO of conformal dimension $\Delta=\frac I2\big|_{I=2}=1$ while the pseudoscalar field $T$ with $M^2_{T}=\frac{(I-3)(I+3)}{L^2}\big|_{I=1}$ is dual to gauge invariant operator with conformal dimension $\Delta= \frac{I+3}2\big|_{I=1}=2$. Then, the gauge/gravity dictionary implies\footnote{The gauge/gravity duality dictionary for the scalar mode is following: Under the asymptotic expansion of a scalar field, 
\begin{align*}
\phi(z,x_{i})
=
\phi_{1}(x_{i})z
+\cdots
+\phi_{d-\Delta}(x_{i})z^{d-\Delta}
+\cdots
+\phi_{\Delta}(x_{i})z^{\Delta}
+\cdots,
\end{align*} the $vev$ of the dual gauge invariant operator with conformal dimension $\Delta$ is proportional to $\phi_{\Delta}(x_{i})$ and the $source$ of that operator is proportional to $\phi_{d-\Delta}(x_{i})$.} the expansion of these two fields in powers of the holographic coordinate $z$ should read
\begin{align}\label{asypExp}
\Psi(z)=V_{\Psi}z+S_{\Psi}z^2+\cdots,\quad T(z)=S_{T}z+V_{T}z^2+\cdots,
\end{align}
where the coefficients $V_{\Psi,T}$  are determined by the $vevs$ of the dual operators and the coefficients $S_{\Psi,T}$ are determined by the values of the external sources which are coupled to those operators.  Comparing \eqref{asypExp} with our results in \eqref{Hmn}, we see that $S_{\Psi}=V_{T}=0$, while $V_{\Psi}=-24\beta_3\mu_0$ and $S_{T}=16\sqrt3\mu_0$. 
Therefore, one can identity the coefficients of $z$ in $\Psi(z)$  as {\it $vevs$} of the CPO with $\Delta = 1$ and the coefficients of $z$ in $T(z)$ as the {\it source} of a gauge invariant operator with $\Delta = 2$ in 3-dimensions. The coupling of these two scalars to gravity in 4-dimensions causes the deformation of the induced metric in \eqref{indH}. More precisely, the $\beta_3$ terms in \eqref{indH} are due to the coupling with $\Psi$, while the numerical terms are due to coupling with $T$. As we discussed in the previous subsection, the matric deformation determines the variation of the HEE ($\delta S_A$), which means that the $\beta_3$-term in $\delta S_A$ in \eqref{4dHEE} is originated from the {\it $vevs$} of the CPO, $\langle{\cal O}^{(1)}\rangle_m$, while the numerical term in $\delta S_A$ is originated from the {\it source} of the gauge invariant operator $J_{\tilde {\cal O}^{(2)}}$.  The variation of HEE is actually related to the squares of the {\it $vevs$} and  the {\it source} term.  
Therefore,  up to $\mu_0^2$-order in the large $N$ limit, the HEE  can be written as
 \begin{align}\label{deltaSA}
\delta S_A = -\frac{4\sqrt{2} \pi\, N^2  l^2}{\sqrt{\lambda}}\left[\frac1{9}\left(J_{\tilde {\cal O}^{(2)}}\right)^2 
+ \frac1{16} \left(\frac{\pi\sqrt{\lambda}\langle {\cal O}^{(1)}\rangle_m}{N^2}\right)^2\right],
\end{align}
where we set the source of the operator $\tilde {\cal O}^{(2)}$ as $J_{\tilde {\cal O}^{(2)}} =\mu_0$.

The CPO of conformal dimension one is as in \eqref{CPO}, and the gauge invariant operator $\tilde {\cal O}^{(2)}$ is built from the fermionic fields of the ABJM theory, $(\psi_{A},$ $A=1,2,3,4)$, and is given by \cite{Bak:2010ry}
\begin{align}\label{CPOt}
\tilde{\cal O}^{(2)}=\tilde C_A^B{\rm Tr}\big(\psi^{\dagger A}\psi_{B}\big),
\end{align}
where $\tilde C_A^B$ are traceless. See \cite{Bena:2000zb} for the source of the mass deformation by adding fermion mass terms in the dual field theory. In general the CPOs of conformal dimension $\Delta$ are given by
\begin{align}\label{CPOD}
{\cal O}^{(\Delta)}=C_{A_1,\cdots,A_{\Delta}}^{B_1,\cdots,B_{\Delta}}{\rm Tr}\big(Y^{A_1}Y^\dagger_{B_1}\cdots Y^{A_{\Delta}}Y^\dagger_{B_{\Delta}}\big).
\end{align}
The gauge invariant operators $\tilde{\cal O}^{(\Delta)}$ are  descendents of the CPOs and they are given by 
\begin{align}\label{CPOtD}
\tilde{\cal O}^{(\Delta)}=\tilde C_{a,A_{1},\cdots,A_{\Delta-2}}^{b,B_1,\cdots,B_{\Delta-2}}{\rm Tr}\big(\psi^{\dagger a}\psi_{b}Y^{A_1}Y^\dagger_{B_1}\cdots Y^{A_{\Delta-2}}Y^\dagger_{B_{\Delta-2}}\big).
\end{align}
The CPOs ${\cal O}^{(\Delta)}$ are dual to the scalar KK modes $\Psi^{I}$ $(I=2,4,6,\cdots)$ with $\Delta=\frac I2$, and the gauge invariant operators $\tilde{\cal O}^{(\Delta)}$ are dual to the pseudoscalar KK modes $T^I$ $(I=1,3,5,\cdots)$ with $\Delta=\frac {I+3}2$.  
We obtain the expansions in holographic coordinate, for those two KK towers of  dual scalars from the asymptotic expansion of the LLM solutions, which are given by
\begin{align}\label{astExp-2}
&\Psi^{2+4i}(z)=\tilde\psi_1(\mu_0z)+\tilde\psi_3 (\mu_0z)^3+\cdots+\tilde\psi_{\Delta} (\mu_0z)^{\Delta}+\tilde\psi_{\Delta+2} (\mu_0z)^{\Delta+2}+\cdots,\quad(\Delta=1+2i),\nn\\
& \Psi^{4+4i}(z)=\tilde\psi_2(\mu_0z)^2+\tilde\psi_4 (\mu_0z)^4+\cdots+\tilde\psi_{\Delta} (\mu_0z)^{\Delta}+\tilde\psi_{\Delta+2} (\mu_0z)^{\Delta+2}+\cdots,\quad(\Delta=2+2i),\nn\\
&T^{1+4i}(z)=t_1(\mu_0z)+t_3 (\mu_0z)^3+\cdots+t_{\Delta-1} (\mu_0z)^{\Delta-1}+t_{\Delta+1} (\mu_0z)^{\Delta+1}+\cdots,\quad(\Delta=2+2i),\nn\\
&T^{3+4i}(z)=t_2(\mu_0z)^2+t_4 (\mu_0z)^4+\cdots+t_{\Delta-1} (\mu_0z)^{\Delta-1}+t_{\Delta+1} (\mu_0z)^{\Delta+1}+\cdots,\quad(\Delta=3+2i),
\end{align}
where $i=0,1,2,\cdots$. As  mentioned before, in these expansions the coefficient of $z^{\Delta}$ is determined by the $vev$ of the dual operator while the coefficient of $z^{d-\Delta}|_{d=3}=z^{3-\Delta}$ is determined by the value of the external source which is coupled to the operator.  Among the towers of the KK modes in \eqref{astExp-2}, the only mode which has nonvanishing coefficient of $z^{3-\Delta}$ is $T=T^{1+4i}\big|_{i=0}$. Therefore, the dual operator $\tilde{\cal O}^{(2)}$ is the only operator which is coupled to an external source, while the rest have no external source term. In addition, from  \eqref{astExp-2}, we see that all the gauge invariant operators dual to the  pseudo-scalar fields $T^{I}$ have vanishing $vev$s whereas all the CPOs which are dual to the scalar fields $\Psi^{I}$ have non-vanishing $vev$s.
This suggests that, the variation of the HEE ($\delta S_A$) can depend on the source term of the gauge invariant operator $\tilde{\cal O}^{(2)}$ and the $vevs$ of the CPOs ${\cal O}^{(\Delta)}(\Delta=1,2,3\cdots)$. However, as can be seen from \eqref{astExp-2}, for $\Delta\ge2$, the $vevs$ are at least quadratic in $\mu_0$, and cannot contribute to $\delta S_A$ at quadratic order in $\mu_0$. Therefore, at quadratic order in $\mu_0$, the variation of HEE is fully determined by the source term of the operator $\tilde{\cal O}^{(2)}$ and the $vev$ of the CPO ${\cal O}^{(1)}$, which is the result in \eqref{deltaSA}.

\subsection{Comments on relative entropy and Fisher information}
The variation of the EE, $\delta S_A$ in \eqref{4dHEE}, is connected with the relative entropy, which is defined as 
\begin{align}
S(\rho || \sigma) \equiv {\rm tr}\left(\rho \log\rho\right) - {\rm tr}\left(\rho \log\sigma\right),
\end{align}
where $\rho$ is a deformed density matrix from some reference density matrix $\sigma$. For the ball-shaped region $A$, the relative entropy is represented as
\begin{align}\label{RelE}
S(\rho_A || \sigma_A) = \Delta \langle H_A\rangle - \Delta S_A, 
\end{align}
where $H_A$ is the modular Hamiltonian associated with the region $A$. See below for details.  
At first order in the deformation parameter, the relative entropy vanishes and \eqref{RelE} becomes the first law of the EE. Therefore, the leading nonvanishing contribution to relative entropy is quadratic in the deformation parameter. In our case the deformation parameter is $\mu_0$, and due to the supersymmetry of the mABJM theory, $\Delta \langle H_A\rangle_{\mu_0^2} =0$ (see the discussion below the equation \eqref{delS1}).  Then using the result in \eqref{4dHEE}, we have 
\begin{align}\label{relSA}
S(\rho_A || \sigma_A)|_{\mu_0^2} = \frac{\pi L^2}{6 G_N^{(4)}}\left( 1 + \frac{\beta_3^2}{32}\right)\left(\mu_0 l\right)^2.  
\end{align}
The relative entropy is positive definite as a measure of {\it distance} between two quantum states and monotonically increasing with the size of the subsystem~\cite{Blanco:2013joa, Lin:2014hva, Lashkari:2014kda, Lashkari:2015hha, Jafferis:2015del,Lin:2016fua}. The relation in \eqref{relSA} represents those properties clearly. The positivity of the relative entropy states the positivity of the Fisher information metric with one deformation parameter,
\begin{align}
F_{\mu_0\mu_0} = \frac{d^2 }{d\mu_0^2}S(\rho_A || \sigma_A)  = \frac{\pi L^2 l^2}{3 G_N^{(4)}}\left( 1 + \frac{\beta_3^2}{32}\right).
\end{align}
This quantity is also known as the fidelity susceptibility. 
In \cite{Lashkari:2015hha}, it was shown that the Fisher information of the reduced density matrix of a ball-shaped subregion at the CFT vacuum is connected to the canonical energy~\cite{Hollands:2012sf} for perturbations in the corresponding Rindler wedge of the AdS geometry. See also \cite{Hyun:2016dvt} for the canonical energy by using the Euler-Lagrange expression and its connection to the Fisher information.   In most cases, for instance \cite{Nozaki:2013vta, Lin:2014hva, Lashkari:2015hha, Beach:2016ocq,Banerjee:2017qti}, the second order deformation in the gravity is connected to  nonvanishing $vev$ of gauge invariant operator in QFT. Though the field fluctuations in gravity side in our case are originated from the $vev$ and $source$ of gauge invariant operators, we expect that the interpretation of the Fisher information in QFT as the canonical energy in the gravity side is correct since the roles of the dual fields $\Psi$ and $T$ in gravity side are indistinguishable.

\section{Gravity from Entanglement and RG Flow}

In the previous  section, we have determined the HEE by applying the RT formula in 4-dimensional gravity, which is obtained from the KK reduction of the 11-dimensional gravity on the LLM geometry. At the quadratic order in $\mu_0$,  our  result is in  complete agreement with the one obtained by applying the RT formula in the 11-dimensional supergravity before the KK reduction~\cite{Kim:2016dzw}. 
This indirectly proves that the solution of the 4-dimensional gravity theory we have built contains all the information of the 11-dimensional LLM geometry near the UV fixed point. In the dual gauge theory, the asymptotic limit of the LLM geometry describes the RG flow from the UV fixed point where the ABJM theory lives.
In this section we discuss the first law-like relation for the EE when there is RG flow due to relevant perturbations from the UV fixed point.

\subsection{Emergent gravity from relevant perturbations in CFT}\label{relpert}

We consider an Euclidean CFT action with a relevant deformation, 
\begin{align}\label{defAct}
I = I^{(0)} + \tilde\lambda \int d^d w \,{\cal O}^{(\Delta)},
\end{align}
where $I^{(0)}$ is the $d$-dimensional CFT action, ${\cal O}^{(\Delta)}$ is a gauge invariant operator with conformal dimension $\Delta$, and $\tilde\lambda$ is the deformation parameter with mass dimension $d-\Delta>0$. In QFT, the EE  is defined as 
\begin{align}\label{SEE}
S_{A} = - {\rm tr} \left(\rho_A \ln \rho_A\right),
\end{align}
 where the total space of states is divided into two subregions $A$, $B$  and  $\rho_A$ is the reduced density matrix of the subregion $A$ at a given time.  Here we consider the subregion $A$ is in a shape of $(d-1)$-dimensional ball of radius $l$.   Under the relevant deformation \eqref{defAct}, the density matrix is deformed as $\rho_A = \rho_A^{(0)} + \delta\rho_A$ with the matrix $\rho_A^{(0)}$ is for the undeformed CFT.  
 
 The EE is calculated  in the path integral formalism using the perturbative expansion in the deformation parameter $\tilde\lambda$.  At the linear order in $\tilde\lambda$, the perturbative evaluation of the EE produces the relation $
\delta S = \delta \langle H_A\rangle_{\tilde\lambda}$, where $H_A$ is the modular Hamiltonian defined as $H_A\equiv -\ln \rho_A $.  This relation is known as the first law of the EE~\cite{Blanco:2013joa,Bhattacharya:2012mi}. For a ball-shaped subregion $A$, the modular Hamiltonian is expressed in terms of energy-momentum tensor  (see below). At the quadratic order of the perturbative expansion, it was pointed out that the variation of EE gets contributions from the two point function $\langle {\cal O}^{(\Delta)} {\cal O}^{(\Delta)}\rangle$  and the three point function $\langle H_A {\cal O}^{(\Delta)} {\cal O}^{(\Delta)}\rangle$~\cite{Rosenhaus:2014woa,Rosenhaus:2014zza}. Later, it was found that in the evaluation of the two point function $\langle {\cal O}^{(\Delta)} {\cal O}^{(\Delta)}\rangle$ there exists an additional finite contribution at the $\tilde\lambda^2$-order, which comes from the non-commutative property between the matrix representations of $\rho_A^{(0)}$ and $\delta\rho_A$~\cite{Faulkner:2014jva}. As we will see later, this finite term reflects the deviation of EE away from the UV fixed point under the relevant deformation and was identified quantitatively in the holographic picture via the RT conjecture.

Next, we follow~\cite{Faulkner:2014jva} and briefly discuss the two terms contributing to the variation of EE at $\tilde\lambda^2$-order.  The first one is given by 
\begin{align}\label{delS1}
\delta S^{(1)} = \frac{\tilde\lambda^2}2 \int d^dw\int d^dw' \langle H_A {\cal O}^{(\Delta)}(w){\cal O}^{(\Delta)}(w')\rangle_0 = \delta\langle H_A\rangle_{\tilde\lambda^2},
\end{align}
where $\langle\cdots\rangle_0$ denotes the $n$-point functions of the undeformed theory. The result in \eqref{delS1} has also been obtained in~\cite{Rosenhaus:2014woa,Rosenhaus:2014zza}. 
For the ball-shaped region $A$, the explicit form of the modular Hamiltonian $H_A$ for the $d$-dimensional CFT is obtained in~\cite{Casini:2011kv}, 
\begin{align}\label{H_A}
H_A = 2\pi \int_{B(l,\vec w_0)} d\Sigma^i \zeta^j T_{ij}= 2\pi \int_{B(l,\vec w_0)} d^{d-1} w \frac{l^2 - |\vec w - \vec w_0|^2}{ 2 l}\, T_{tt}(t_0,\vec w),
\end{align} 
where $i =0,1,\cdots,d-1$,  the ball $B(l,\vec w_0)$ is on a time slice $t=t_0$, it is of radius $l$  and centered at $\vec w = \vec w_0$. The $d\Sigma^i$ is the volume form on the $(d-1)$-dimensional surface perpendicular to a unit vector in $i$-th direction and $\zeta$ is the conformal Killing vector defined as  
\begin{align}\label{Kzeta}
\zeta = \left(\frac{l^2 -\rho^2 - t^2}{2l}\right)\partial_t - \frac{\rho t}{l}\partial_\rho
\end{align}   
with radius $\rho=\sqrt{w_1^2 + \cdots + w_{d-1}^2}$. 
Here $T_{tt}$ denotes the $(tt)$-component of the energy-momentum tensor in $d$-dimensional CFT. 
Inserting \eqref{H_A} into \eqref{delS1}, the calculation of $\delta S^{(1)}$ is reduced to the evaluation of the three point function $\langle T_{tt}(w) {\cal O}^{(\Delta)} (w') {\cal O}^{(\Delta)}(w'')\rangle_0$. 
In the mABJM theory with supersymmetric discrete Higgs vacua,  the three point function is vanishing in the large $N$ limit. In order to see this, we expand the field near the vacua as $Y^{A=1,2,3,4} = Y_0^A + \tilde Y^A$ with the vacuum configuration $Y_0^A$. Then the gauge invariant operators are written as 
\begin{align}
{\cal O}^{(\Delta)}(w) = {\cal O}^{(\Delta)}_0 + \sum_i\tilde {\cal O}_i^{(\Delta)}(w), \label{opDel}
\end{align}
where ${\cal O}^{(\Delta)}_0\equiv {\cal O}^{(\Delta)}(w)|_{Y^A = Y_0^A}$ and $\tilde {\cal O}^{(\Delta)}_i$'s are operators expanded around the vacuum $Y_0^A$. Then the three point function can formally be written as 
\begin{align}
\langle T_{tt}(w) {\cal O}^{(\Delta)} (w') {\cal O}^{(\Delta)}(w'')\rangle_0 &=
\sum_{i,j} \langle T_{tt}(w) \tilde{\cal O}_i^{(\Delta)} (w') \tilde{\cal O}_j^{(\Delta)}(w'')\rangle_0+\left({\cal O}_0^{(\Delta)}\right)^2\langle T_{tt}(w)\rangle_0\nn\\
&+{\cal O}_0^{(\Delta)}\sum_i\left[\langle T_{tt}(w)\tilde {\cal O}_i^{(\Delta)}(w')\rangle_0+\langle T_{tt}(w)\tilde {\cal O}_i^{(\Delta)}(w'')\rangle_0\right]. 
\end{align}
In CFT, the conformal invariance dictates that the two point function $\langle {\cal O} T_{\mu\nu}\rangle_0$ is vanishing. In addition, the first term on the right-hand side in the above equation is a multi-trace term and is suppressed by $1/N$ as compared with the single trace terms. Therefore the three point function in the large $N$ limit is given by 
\begin{align}
\langle T_{tt}(w) {\cal O}^{(\Delta)} (w') {\cal O}^{(\Delta)}(w'')\rangle_0 &=
\left({\cal O}_0^{(\Delta)}\right)^2\langle T_{tt}(w)\rangle_0 + \frac1{N}-{\rm corrections}. 
\end{align}
Since the mABJM theory is a supersymmetric gauge theory, the $vev$ of the energy-momentum tensor $\langle T_{tt}(w)\rangle_0$ is vanishing. Therefore, we expect that $\delta S^{(1)}$ for the mABJM theory is vanishing in the large $N$-limit.

The second contribution to the variation of the EE is obtained in \cite{Faulkner:2014jva} and it is given by  
\begin{align}\label{delS2}
\delta S^{(2)} = -2\pi \int_{{\cal H}^+} d\Sigma^\mu\xi^\nu \tilde{T}_{\mu\nu},
\end{align}
where $\mu,\,\nu = 0,1,\cdots d$ including one additional direction, which will be identified in dual gravity theory as the holographic direction.
Here the integration is performed over the region ${\cal H}^+$, which is the future part of the Rindler horizon in the emergent AdS$_{d+1}$ space and $d\Sigma^\mu$ is the surface element on the horizon.  
The energy-momentum tensor $\tilde{T}_{\mu\nu}$ of an auxiliary scalar field $\tilde\phi$, is
\begin{align}\label{TAab}
\tilde{T}_{\mu\nu} = \nabla_\mu\tilde\phi \nabla_\nu\tilde\phi - \frac12 g_{\mu\nu}^{(0)}\left( \nabla_\lambda\phi\nabla^\lambda\tilde\phi + m^2\tilde\phi^2\right),
\end{align} 
which satisfies the conservation law $\nabla_\mu\tilde{T}^{\mu\nu} = 0$, and this means the $\delta S^{(2)}$ in \eqref{delS2} is a conserved charge. The field $\tilde\phi$ satisfies the Klein-Gordon equation, 
\begin{align}\label{phieq}
\nabla^2\tilde\phi - \frac{\Delta (\Delta- d)}{L_{{\rm AdS}}^2}\tilde\phi = 0
\end{align} 
with some boundary conditions at $z=0$.
For calculation of $\delta S^{(2)}$, we consider an explicit example of the emergent space. To do that, we introduce the AdS$_{d+1}$ geometry in the Poincare patch, 
\begin{align}\label{ds2}
ds^2 = \frac1{z^2}\left(-dt^2 + dz^2 + d\rho^2 + \rho^2 d\Omega_{d-2}^2\right), 
\end{align}
where $d\Omega_{d-2}^2$ is the line element on $S^{d-2}$ with unit radius. The Killing vector $\zeta$ in \eqref{Kzeta} of the $d$-dimensional CFT extends to a Killing vector in the emergent bulk geometry,  
\begin{align}\label{Kvecxi}
\xi = \frac{l^2 -z^2-\rho^2-t^2}{2l}\partial_t - \frac{t}{l}\left(z\partial_z + \rho\partial_\rho\right).
\end{align}
When we choose a spatial slice at $t=0$, the AdS Rindler horizon is reduced to the minimal surface satisfying the relation $z(\rho) = \sqrt{l^2-\rho^2}$ in \eqref{solz}. Then the  variation $\delta S^{(2)}$ in \eqref{delS2} is given by 
\begin{align}\label{delS2-3}
\delta S^{(2)} = -\frac{\pi \Omega_{d-2} L_{{\rm AdS}}^{d-1}}{l}\int_{z_\Lambda}^l dz z^{1-d}\int_0^{\sqrt{l^2-z^2}}d\rho \rho^{d-2}\left(l^2-z^2-\rho^2\right)\tilde T_{tt},
\end{align}
where  $\Omega_{d-2}= \frac{2\pi^{(d-1)/2}}{\Gamma\left((d-1)/2\right)}$. We used the volume form $d\Sigma^t = L_{{\rm AdS}}^{d-1}z^{1-d}dz \rho^{d-2} d\rho d\Omega_{d-2}$ and the Killing vector $\xi^t = \frac{l^2-z^2-\rho^2}{2l}$, and  also introduced the cutoff scale $z_\Lambda$ to regularize the divergence at $z=0$.

In the $z\to 0$ limit, the $\tilde\phi$ configuration satisfying the equation \eqref{phieq} is given by 
\begin{align}\label{phisol}
\tilde\phi(z, w)\longrightarrow V_{\tilde\lambda}( w) z^\Delta + S_{\tilde\lambda}( w) z^{d-\Delta}.
\end{align}
Since this auxiliary field is absent in the undeformed theory, the coefficients $V_{\tilde\lambda}$ and $S_{\tilde\lambda}$ also depend on the  deformation parameter\footnote{If one identifies this auxiliary field with a scalar field in a gravity theory, the gauge/gravity dictionary dictates that $V_{\tilde\lambda}$ corresponds to the $vev$ of a gauge invariant operator with conformal dimension $\Delta$, while $S_{\tilde\lambda}$ corresponds to the source coupled to the operator.}.
Inserting the solution \eqref{phisol} into \eqref{delS2-3}, one can obtain $\delta S^{(2)}$ in the path integral method. 
For a very small ball-shaped region, $l\ll L_{{\rm AdS}}$, the coefficients $V_{\tilde\lambda}$ and $S_{\tilde\lambda}$ can be regarded as constants,
 which are consistent with our case.
In the literature the $\Delta<\frac d2$ and $\Delta>\frac d2$ cases are treated separately, whereas the $\Delta=\frac d2$ case needs a special  treatment~\cite{Casini:2016rwj,Speranza:2016jwt}.

{\bf $\bullet$ $\Delta < \frac{d}{2}$ case:}
 To obtain the leading contribution for the small value of $\tilde\lambda$ in $\delta S^{(2)}$, it is enough to consider the  asymptotic behavior of $\tilde\phi(z,w)$ as 
\begin{align}\label{glam=0}
\tilde\phi(z,w) = V_{\tilde\lambda} z^\Delta, 
\end{align}
where $V_{\tilde\lambda} = \mathbb{N}_V\tilde\lambda$ with a numerical factor $\mathbb{N}_V$. Then we obtain the $(tt)$-component of the energy-momentum tensor, $T_{tt}^A = V_{\tilde\lambda}^2 z^{2\Delta -2} \Delta \left(\Delta - \frac{d}{2}\right)$. 
Inserting this into \eqref{delS2-3}, we obtain~\cite{Blanco:2013joa,Speranza:2016jwt} 
\begin{align}\label{delS2-1}
\delta S^{(2)} &=  \frac{\pi \Omega_{d-2} \mathbb{N}_V^2\tilde\lambda^2L_{{\rm AdS}}^{d-1}}{l}\Delta\left(\frac{d}{2} - \Delta\right)\int_{z_\Lambda}^{l} z^{2\Delta - d -1} \int_0^{\sqrt{l^2- z^2}} d\rho \rho^{d-2}\left(l^2 - z^2- \rho^2\right)
\nn \\
&= -\frac{\pi \Omega_{d-2} \mathbb{N}_V^2\tilde\lambda^2L_{{\rm AdS}}^{d-1} l^{2\Delta}\Delta \Gamma\left(\frac{d+3}{2}\right)\Gamma\left(\Delta-\frac{d}{2}+1\right)}{(d^2-1) \Gamma\left(\frac32 + \Delta\right)}+ \delta S^{(2)}_{{\rm div}},
\end{align}
where $\delta S^{(2)}_{{\rm div}}$ includes the divergent pieces depending the UV-cutoff $z_\Lambda$, 
\begin{align}
\delta S^{(2)}_{{\rm div}}= -\frac{\pi \Omega_{d-2} \mathbb{N}_V^2\tilde\lambda^2L_{{\rm AdS}}^{d-1} l^{d}\Delta}{(d^2-1)z_\Lambda^{d-2\Delta}}+ {\cal O}\left(z_\Lambda^{-d + 2\Lambda + 2}\right).
\end{align}

{\bf $\bullet$ $\Delta > \frac{d}{2}$ case:}
In order to extract the leading contribution for the small value of $\tilde\lambda$, we consider the asymptotic behavior of $\tilde\phi(z,w)$ as 
\begin{align}\label{flam=0}
\tilde\phi(z,w) = S_{\tilde\lambda} z^{d-\Delta}, 
\end{align}
where $S_{\tilde\lambda} = \mathbb{N}_S \tilde\lambda$ with a numerical factor $\mathbb{N}_S$. Then we obtain the $(tt)$-component of the energy-momentum tensor, $T_{tt}^A = S_{\tilde\lambda}^2 z^{2d -2\Delta -2} \left(\Delta - \frac{d}{2}\right) \left(\Delta -d\right)$. 
Inserting this into \eqref{delS2-3}, we obtain the result in \cite{Faulkner:2014jva},
\begin{align}\label{delS2-2}
\delta S^{(2)} &=  \frac{\pi \Omega_{d-2} \mathbb{N}_S^2\tilde\lambda^2L_{{\rm AdS}}^{d-1}}{l}(d-\Delta)\left(\Delta-\frac{d}{2} \right)\int_{z_\Lambda}^{l} z^{d-2\Delta  -1} \int_0^{\sqrt{l^2- z^2}} d\rho \rho^{d-2}\left(l^2 - z^2- \rho^2\right)
\nn \\
&= -\frac{\pi \Omega_{d-2} \mathbb{N}_S^2\tilde\lambda^2L_{{\rm AdS}}^{d-1} l^{2(d-\Delta)}(d-\Delta) \Gamma\left(\frac{d+3}{2}\right)\Gamma\left(\frac{d}{2}-\Delta +1\right)}{(d^2-1) \Gamma\left(\frac32 +d- \Delta\right)}+ \delta S^{(2)}_{{\rm div}},
\end{align}
where 
\begin{align}
\delta S^{(2)}_{{\rm div}}= -\frac{\pi \Omega_{d-2} \mathbb{N}_S^2\tilde\lambda^2L_{{\rm AdS}}^{d-1} l^{d}(d-\Delta)}{(d^2-1)  z_\Lambda^{2\Delta-d}}+ {\cal O}\left(z_\Lambda^{d - 2\Lambda + 2}\right).
\end{align}
The leading finite contribution in \eqref{delS2-2} was also obtained in \cite{Liu:2012eea} using the RT formula. 

In the calculations of $\delta S^{(1)}$ in \eqref{delS1} and $\delta S^{(2)}$ in \eqref{delS2}, there exist divergent terms. 
One needs to subtract those terms by adding an appropriate count term in $\tilde\lambda^2$-order
\begin{align}\label{deltaS_A} 
\delta S_A =  \langle H_A\rangle_{\tilde\lambda} -2\pi \int_{{\cal H}^+} d\Sigma^\mu\xi^\nu \tilde T_{\mu\nu} + S_{ct}.
\end{align}

\subsection{Gravity from entanglement in the mABJM theory}

In the previous subsection, using the path integral method in $d$-dimensional CFT in the presence of the relevant deformation \eqref{defAct}, we have summarized that some part of $\delta S$ in $\tilde\lambda^2$-order becomes a conserved charge in $(d+1)$-dimensional theory.
The conserved charge is defined by introducing one additional coordinate $z$ and one auxiliary field $\tilde \phi$.
The appearance of the additional coordinate indicates the emergence of gravity from the entanglement in QFT. We expect the emergent gravity identifies with the gravity theory which is dual to the QFT.  In particular, the energy-momentum tensor of the auxiliary field is expected to be identified with the energy-momentum tensor of a dynamical scalar field in the dual gravity theory. 

In this paper, we have constructed a 4-dimensional gravity theory with the matter sector composed of two scalar fields $\Psi$ and $T$.  The result we summarized in the previous subsection suggests that these two scalar fields should emerge from the calculation of the variation of the EE in the 3-dimensional dual mABJM theory.    In this subsection, we discuss this phenomena in detail and show the emergence of 4-dimensional Einstein equation from the EE analysis.

In order to calculate the quantity $\delta S^{(2)}$ in \eqref{delS2} from the energy-momentum tensor defined in \eqref{Tmn}, we treat the contributions from $\Psi$ and $T$ separately. The operator which is dual to the field $\Psi$ is of conformal dimension one and it corresponds to the case $\Delta <\frac{d}{2}$, while that of the field $T$ is of conformal dimension two and corresponds to the case $\Delta >\frac{d}{2}$. Comparing the energy-momentum tensors in \eqref{Tmn} and \eqref{TAab}, we need to rescale the scalar fields as $\tilde \Psi = \sqrt{A_\psi} \Psi$ and $\tilde T = \sqrt{A_t} T$ and then asymptotic behaviours of the scalar fields in \eqref{Hmn} are rescaled as 
\begin{align}\label{tilPsi}
\tilde \Psi = -24\sqrt{A_\psi}\,\beta_3\mu_0 z + {\cal O}(\mu_0^3),
\qquad 
\tilde T = 16\sqrt{3}\sqrt{A_t}\,\mu_0 z + {\cal O}(\mu_0^3).
\end{align} 
The energy-momentum tensors for $\tilde \Psi$ and $\tilde T$ up to $\mu_0^2$-order are read from \eqref{TAab}, 
\begin{align}\label{tilTij}
\tilde T_{ij}^{(\tilde \Psi)} &= 288 A_\psi\beta_3^2\mu_0^2\eta_{ij},\qquad 
\tilde T_{zz}^{(\tilde \Psi)} = 864 A_\psi\beta_3^2\mu_0^2,
\nn \\
\tilde T_{ij}^{(\tilde T)} &= 384 A_t\mu_0^2,\qquad 
\tilde T_{zz}^{(\tilde T)} = 1152 A_t\mu_0^2.
\end{align}

For the field $\tilde\Psi$ in \eqref{tilPsi}, we obtain the variation of the EE from  \eqref{delS2-1} after the cancellation of the divergent term,
\begin{align}
\delta S^{(2)}_\Psi =  -96 \pi^2 L^2  A_\psi \beta_3^2 (\mu_0 l)^2.
\end{align}
Similarly, for the scalar field $\tilde T$, the result is obtained from \eqref{delS2-2},
\begin{align}
\delta S^{(2)}_T =  -128 \pi^2 L^2  A_t  (\mu_0 l)^2.
\end{align} 
Therefore, the total variation of the EE is given by 
\begin{align}\label{delS2-4}
\delta S^{(2)} = \delta S^{(2)}_\Psi+\delta S^{(2)}_T = -128\pi^2 L^2   \left(A_t+ \frac{3 A_\psi \beta_3^2}{4}  \right)(\mu_0 l)^2.
\end{align}

In the subsection \ref{relpert}, we argued that the quantity $\delta S^{(1)}$ for the mABJM theory is vanishing in the large $N$ limit. This is because of vanishing vacuum energy density $\langle T_{ij}(w)\rangle_0 = 0$ in supersymmetric theories. Now we examine again the vanishing of $\delta S^{(1)}$ for the mABJM theory using the gauge/gravity duality dictionary.
To that end,  we start from the deformed 4-dimensional  metric in \eqref{Hmn} and \eqref{4Dgmn},  
\begin{align}\label{oldFG}
ds^2 = \frac{L^2}{4z^2}\left[\left(1- {\cal B}(\mu_0z)^2+{\cal O}( z^4)\right)dz^2+\left(1- {\cal A}(\mu_0z)^2+{\cal O}( z^4)\right)\eta_{ij} dx^i dx^j  \right],
\end{align}
where 
\begin{align}
{\cal A} =\frac1{45}\left(30 + \beta_3^2\right), 
\qquad {\cal B} = \frac1{360}\left(960 + 29\beta_3^2\right). 
\end{align}
The FG coordinate system is convenient to read the $vev$ of the energy-momentum tensor from the asymptotic expansion of the metric. Therefore, we apply the coordinate transformation, 
\begin{align}
z\,\longrightarrow\, \tilde z = z - \frac{\mu_0^2 {\cal B}}{4} z^3, 
\end{align}
to the metric \eqref{oldFG} in order to write it in FG-coordinate system, 
\begin{align}\label{newFG}
ds^2 = \frac{L^2}{4\tilde z^2}\left[ d\tilde z^2 + \left(1- \frac12\left(2 {\cal A} + {\cal B}\right)(\mu_0 \tilde z)^2 + {\cal O}(\tilde z^4)\right)\eta_{ij}dx^i dx^j\right].
\end{align}
This asymptotic behavour of the metric tells us the fact that $\langle T_{ij}\rangle_0$ is vanishing\footnote{For the metric in the FG coordinate system in ($d+1$)-dimensional gravity theory, 
\begin{align}
ds^2 = \frac{L_{\rm AdS}^2}{z^2}\left(dz^2 + g_{ij}(x,z) dx^idx^j\right),
\end{align}
where $L_{\rm AdS}$ is the radius of the AdS$_{d+1}$ geometry and the  metric is expanded as $g_{ij}(x,z) = g_{(0)ij}(x) + z^2 g_{(2)ij}(x) +\cdots + z^d g_{(d)ij}(x)+\cdots$ in the asymptotic limit, the $vev$ of the energy-momentum tensor operator is given by~\cite{Balasubramanian:1999re,deHaro:2000vlm,Skenderis:2000in,Bianchi:2001kw} 
\begin{align}\label{vevTij}
\langle T_{ij}\rangle = \frac{d L_{{\rm AdS}}^{d-1}}{16\pi G_N}\, g_{(d)ij}. 
\end{align}}.  
This confirms the claim that $\delta S^{(1)} =0$ for the mABJM theory.

In conclusion, the quantity $\delta S^{(2)}$ represents the full variation of the EE in the mABJM theory, which coincides with the variation of area obtained in \eqref{gammaA1}.  On the other hand, for a general metric perturbation, the variation of the area is also given~\cite{Wald:1993nt,Faulkner:2013ica}, 
\begin{align}\label{gammaA3}
\delta \gamma_A = -\int_{{\cal H}_0} d\Sigma^\mu\xi^\nu \delta G_{\mu\nu} + \delta \gamma_A^{(ct)}, 
\end{align}
where ${\cal H}_0$ is a region on the space-like surface $t=0$ squeezed between the minimum area and the disk $A$ with the UV cut-off at $z=z_\Lambda$. The term $\delta \gamma_A^{(ct)}$ is introduced to cancel out the divergences which arise from $z_\Lambda\to 0$ limit. 
Here the variation of the Einstein tensor $\delta G_{\mu\nu}$ is read from the left-hand side of \eqref{Hmn-eq}, 
\begin{align}\label{deltaG}
\delta G_{\mu\nu} = &\frac12\Big(-\square   H_{\mu\nu}+\nabla^\rho\nabla_\mu   H_{\rho\nu}+\nabla^\rho\nabla_\nu   H_{\rho\mu}-\nabla_{\mu}\nabla_{\nu}  H\Big)+\frac{12}{L^2}  H_{\mu\nu}-\frac{6}{L^2}g_{\mu\nu}  H
\nn\\
-&\frac12g_{\mu\nu}(\nabla^\rho\nabla^\sigma  H_{\rho\sigma}-\square  H).
\end{align}
Inserting \eqref{Hmn} into \eqref{deltaG}, we obtain 
\begin{align}\label{deltaG2}
\delta G_{ij} = \frac{\mu_0^2}{8}\left(32 + \beta_3^2\right)\eta_{ij},\qquad \delta G_{zz} = \frac{3\mu_0^2}{8} \left(32 + \beta_3^2\right). 
\end{align}
Plugging \eqref{deltaG2} into \eqref{gammaA3}, we obtain 
\begin{align}\label{gammaA4}
 -\int_{{\cal H}_0} d\Sigma^t \xi^t \delta G_{tt} = -\frac{\pi L^2}{48}\left(32 + \beta_3^2\right) (\mu_0 l)^2 + \frac{\pi L^2}{128}\frac{l}{z_\Lambda}\left(32 + \beta_3^2\right) (\mu_0 l)^2.
\end{align}
The divergent term in $z_\Lambda \to 0$ limit in \eqref{gammaA4} is cancelled by $\delta \gamma_A^{(ct)}$ in \eqref{gammaA3} and the finite variation of the area  is equivalent to the $\delta\gamma_A$ in \eqref{gammaA1}. Finally, we identify the variation of the EE obtained from field theory calculations with the one obtained from  the  RT formula, 
\begin{align}\label{tilTG}
-2\pi \int_{{\cal H}^+} d\Sigma^\mu\xi^\nu \tilde T_{\mu\nu} =  -\frac{1}{4G_N^{(4)}}\int_{{\cal H}_0} d\Sigma^\mu\xi^\nu \delta G_{\mu\nu}. 
\end{align}
For the Killing vector $\xi^\mu$, the term in the left-hand side of \eqref{tilTG} defines a conserved quantity and thus one can choose any surface homologous to ${\cal H}^+$ as the integration surface. For the choice of the surface at $t=0$, the integration surfaces of both sides in \eqref{tilTG} are identified.  
Then, from this relation, we obtain 
\begin{align}\label{EinEq}
\delta G_{\mu\nu} = 8\pi G_N^{(4)} \tilde T_{\mu\nu},
\end{align}
which is the linearized Einstein equation with matter fields. 
Inserting \eqref{tilTij} and \eqref{deltaG2} into \eqref{EinEq}, one can determine the numerical factors $A_\psi$ and $A_t$, which exactly match the values given in \eqref{coefs}.  Therefore, one can see that the RT formula satisfying the Einstein equation reproduces the variation of the EE calculated in the path integral method in the field theory side.

\section{Conclusion}

In this paper, we investigated the phenomena of the emergent gravity in 4-dimensions from the EE of the 3-dimensional mABJM theory. Using the path integral method developed in \cite{Faulkner:2014jva} and the RT formula, we clarified the relation between the emergent (auxiliary) gravity and the Einstein-Hilbert action with two scalar fields in 4-dimensions, which is obtained from the KK reduction of the 11-dimensional supergravity on the LLM geometries with ${\mathbb Z}_k$ orbifold.

Our analysis relies heavily on the gauge/gravity duality between the mABJM theory and the 11-dimensional supergravity on the LLM geometries. 
In order to setup the gauge/gravity dictionary, we need to construct the 4-dimensional gravity using the KK reduction on the compact manifold $S^7/{\mathbb Z}_k$. In our previous work~\cite{Jang:2016tbk}, we showed an exact dual relation for the $vev$ of a CPO of conformal dimension one (${\cal O}^{(1)}$) in mABJM theory and a scalar field in an asymptotically AdS$_4$ gravity theory in the large $N$ limit. However, the connection between the 4-dimensional graviton mode and the 11-dimensional fluctuations was missing. In this paper, for the minimal ingredients that encode all the information of the LLM geometries with ${\mathbb Z}_k$ orbifold in the asymptotic limit up to $\mu_0^2$-order, we completed the non-trivial KK maps between the 4-dimensional fields and the 11-dimensional fluctuations on AdS$_4\times S^7/{\mathbb Z}_k$. The resulting 4-dimensional fields are composed of the graviton mode $H_{\mu\nu}$, one scalar field $\Psi$, and one pseudoscalar field $T$.  In the matter sector, the asymptotic behaviour of the scalar field $\Psi$  determines the $vev$ of ${\cal O}^{(1)}$ whereas the  asymptotic behaviour of the pseudoscalar field $T$ determines the $source$ which couples to a gauge invariant operator of conformal dimension two ($\tilde{\cal O}^{(2)}$).

The presence of these scalar fields in the 4-dimensional gravity theory implies the existence of some relevant deformation in the dual CFT and as a result the EE is expected to show a variation from that of the CFT.  Employing the holographic RT formula for subregion $A$, which is a disk of radius $l$, we calculated the leading order contribution to the variation of the EE $\delta S_A$ in mABJM theory by using the 4-dimensional metric ($g_{\mu\nu} +H_{\mu\nu}$), which encodes the information of the asymptotic LLM geometries in small mass limit.  We showed that, the leading order contribution to the $\delta S_A$ is quadratic in the deformation parameter $\mu_0$. 
At such leading order, $\delta S_A$ is completely fixed by the $vev$ of 
${\cal O}^{(1)}$ and the $source$ which couples to $\tilde{\cal O}^{(2)}$. The obtained $\delta S_A$ is negative, which is consistent with the $F$-theorem in 3-dimensional gauge theory and describes the RG flow from UV fixed point of the conformal invariant ABJM theory.

Based on a recent progress in the computation of the EE by using the path integral method in CFT with some relevant deformations, we reproduced the HEE for the mABJM theory. In order to calculate the variation of EE using the path integral method,  it is necessary to introduce an additional coordinate $z$ and one auxiliary scalar field for every relevant operator added to the CFT action. In the quadratic  order of approximation, the mABJM theory is regarded as a deformation of the ABJM theory by the relevant operators ${\cal O}^{(1)}$  and  $\tilde{\cal O}^{(2)}$. Therefore, we need to introduce two auxiliary scalar fields.  We identified these auxiliary fields with the scalar fields $\Psi$ and $T$, in the calculation of the EE using the path integral approach.
Consistent with the holographic method, the leading order contribution to $\delta S_A$ from the path integral approach is quadratic in the deformation parameter.  
Furthermore, the variation of HEE obtained from the RT formula and the $\delta S_A$ from the path integral methods are equal, when the linearized Einstein equation with the energy-momentum tensor of the two scalar fields $\Psi$ and $T$, is satisfied.   
This Einstein equation agrees  with the one we obtained from the KK reduction of the 11-dimensional supergravity on LLM geometries. This agreement and the appearance of the additional coordinate $z$ are the indications of the  emergence of an asymptotically AdS$_4$ gravity from the quantum entanglement of the 3-dimensional mABJM theory.

In the calculation of $\delta S_A$ in terms of the path integral method developed in \cite{Faulkner:2014jva}, we used the energy-momentum tensor for the two scalar fields $\Psi$ and $T$ in the dual 4-dimensional gravity theory, relying on the exact dual relation in our previous work in \cite{Jang:2016tbk}.  However, in general, one can compute the $\delta S_A$ up to $\mu_0^2$-order in the path integral method by using the mABJM theory directly without using the dual theory. We leave this for future work.


\section*{Acknowledgements}
OK would like to thank the participants of the 5th IBS Brainstrom workshop for stimulating discussions and appreciates APCTP for its hospitality during completion of this work. DT would like to thank the physics department of Addis Ababa University for hospitality, during the visit to present part of this work. This work was supported by the National Research Foundation of Korea(NRF) grant with grant number NRF-2016R1D1A1B03931090 (Y.K.),  NRF-2017R1D1A1A09000951 (O.K.), and NRF-2017R1D1A1B03032523 (D.T.). 

\end{document}